# Spatiotemporally Separable Biphoton State Generated by Spontaneous Four Wave Mixing in Ultrathin Nonlinear Films

Chenzhi Yuan [1, 2], Wei Zhang [1, 2]*, and Yidong Huang [1, 2]

1) Beijing National Research Center for Information Science and Technology (BNRist), Beijing Innovation Center for Future Chips, Frontier Science Center for Quantum Information, Electronic Engineering Department, Tsinghua University, Beijing 100084, China
2) Beijing Academy of Quantum Information Sciences, Beijing 100193, China
*zwei@tsinghua.edu.cn

We theoretically investigated the biphoton state generated by spontaneous four wave mixing (SpFWM) in ultrathin nonlinear films. The expression of the biphoton state is obtained by perturbation method, in which the longitudinal phase mismatch term is eliminated due to the ultra-small nonlinear interactive length. As a result of the relaxation of the longitudinal phase match condition, frequency bandwidth of the biphoton state would be very large. The correlation function of the biphoton state is analyzed, showing that the space and time in the biphoton correlation of this state are factorable. The calculations of the frequency purity further indicate that the temporal and spatial degrees of freedom in this biphoton state are separable. The spatial Schmidt numbers of the biphoton state are also calculated, showing that this state supports high dimensional transverse entanglement. These results shows that SpFWM in ultrathin nonlinear films is promising in generating biphoton states for applications involving hyper-entanglement and high-dimensional entanglement.

## I. Introduction

Spontaneous parametric nonlinear optical processes, including spontaneous parametric down conversion (SPDC) [1, 2] and spontaneous four wave mixing (SpFWM) [3-5], are important ways to generate correlated biphoton states, which have been widely investigated and applied in various experiments of quantum information processing [6] and quantum communications [7, 8] involving photons. In these processes, the energy and momentum conservations between the annihilated and generated photons determine the properties of quantum correlation in the biphoton states [9-19]. The momentum conservation can be expressed by the phase matching conditions, which is largely determined by the geometrical structure of the nonlinear media.

Usually, nonlinear optical bulk crystals [13] or periodically poled waveguides [20] with length of ~mm are employed in SPDC. Due to the considerable interaction length, the longitudinal phase mismatch becomes remarkable and brings some limits to the generated biphoton states. For example, in these states the temporal and spatial degrees of freedom are not separable, limiting their applications in realization of hyper-entanglements [21]. It also leads to the difficulty on the definitions of temporal and spatial coherences in these states. Instead, the coherence of these states is defined in an "X" trajectory in the spatial and temporal dimensions [10, 15-16]. On the other hand, the longitudinal phase mismatch also limits the frequency and spatial frequency bandwidths of the biphoton states [12], which are important for quantum metrology techniques based on biphoton interferences [22, 23]. For the biphoton state generation by SpFWM, usually the third order nonlinear waveguides, such as optical fibers [3] and silicon waveguides [4], are employed. Long waveguide lengths are required (several hundreds meters for optical fibers and several millimeters for silicon waveguides) to compensate the low third order nonlinearity in these materials. Hence, the longitudinal phase mismatch is also important. It mainly impacts on the frequency bandwidth of the biphoton states if the processes are stimulated only in a specific waveguide mode, or among several known modes [24].

In recent years, nonlinear optical effects in nonlinear thin films attracted much attention. The thicknesses of these nonlinear films are very small, which relax the requirement of longitudinal phase matching. It has been demonstrated by four wave mixing in thin metallic films and graphite thin films [25, 26]. It also was proposed that the longitudinal phase matching factor can be eliminated in biphoton states generated by SPDC in thin films [27]. Recently, C. Okoth, *et. al.* reported an SPDC experiment employing $LiNbO_3$ with thickness of 5.8~6$\mu$m, showing that photon pairs with extremely broad frequency spectrum and wide emission direction were obtained due to the relaxation of phase matching condition [28]. However, the theoretical analysis on the biphoton states generated in ultra-thin nonlinear films is still difficult, which limits the investigation on the properties of the states, such as its correlation function, spatiotemporal separability and high dimensional transverse entanglement. Ref. [29] pointed out that the spatial mode analysis of biphoton states generated by SPDC would have problems of divergence if the phase matching factor disappears. It can be expected that similar problems also exists in analyzing other properties of the biphoton states generated by spontaneous parametric optical processes in ultrathin films, such as correlation function and transverse entanglement.

In this work, we theoretically investigated the properties of biphoton states generated by SpFWM in ultrathin nonlinear films comprehensively. The expression of the biphoton state is obtained by perturbation method, showing that the ultrathin film structure relaxes the requirement of phase matching in the SpFWM process. Then, the correlation function of the biphoton state is deduced, showing that the relaxation of phase match leads to the time-space factorability in the correlation function. It also leads to the spatiotemporal separatability of the biphoton state, which is analyzed by the frequency purity when the spatial and polarization degrees of freedom are traced out. On the other hand, the spatial Schmidt numbers are calculated to show that high-dimensional transverse entanglement can be realized by this state. Finally, the feasibility of experimental observation of this biphoton state is evaluated numerically using different nonlinear ultrathin materials.

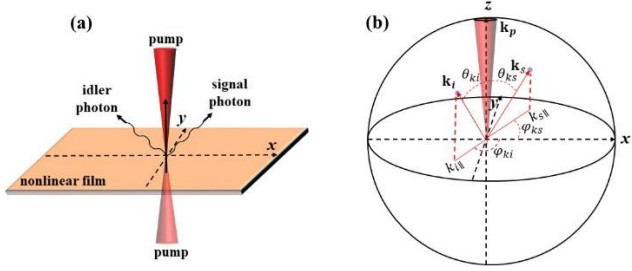

Fig.1 The illustrations of SpFWM in an ultrathin film. (a) The sketch shows the focused pump beam (red cone) and signal/idler photon emissions (wave lines). (b) The geometrical configuration in **k**-space.

## II. EXPRESSION OF BIPHOTON STATES GENERATED BY SpFWM IN ULTRATHIN FILM

The sketch of the biphoton state generation by SpFWM in a ultrathin film is shown in Fig. 1 (a). The film is made by a medium with thickness of $d$ and high third order nonlinearity. A linearly polarized pulsed pump light propagating along positive $z$ direction, which is shown by the red beam, illuminates the film at a spot. The polarization direction of the pump beam is defined as $x$-axis and the surface of the ultrathin film is the $x$-$y$ plane. Due to the ultrathin property of the nonlinear film and the polarization of pump beam, only the components $\chi^{(3)}_{xxxx}$ and $\chi^{(3)}_{yyxx}$ in the nonlinear susceptibility tensor are effective in SpFWM. The signal and idler photons generated by SpFWM would emit from the spot. To make the theoretical derivation simple, we assume that $d$ is small enough to neglect the propagation losses of the pump light, generated signal and idler photons.

To analyze the generated biphoton state, the Schrödinger equation of SpFWM process is solved by perturbation method under low parametric gain [3, 9]. The solution provides the expression of the biphoton state $|\psi\rangle_{s-i}$

$$|\psi\rangle_{s-i} = \sum_{j_s=0,1}\sum_{j_i=0,1}\sum_{l_s=0,1}\sum_{l_i=0,1}\int d^6\mathbf{S}\,\psi(\mathbf{S},j_s,l_s,j_i,l_i)$$
$$\times a_s^\dagger(\mathbf{S}_s,j_s,l_s)a_i^\dagger(\mathbf{S}_i,j_i,l_i)|0\rangle_s|0\rangle_i, \quad (1)$$

where $a_{s,i}^\dagger(\mathbf{S}_{s,i},j_{s,i},l_{s,i})$ ($a_{s,i}(\mathbf{S}_{s,i},j_{s,i},l_{s,i})$) is the creation (annihilation) operator for a mode with indices $\mathbf{S}_{s,i}=[\omega_{s,i},\mathbf{k}_{s,i\parallel}]$, $j_{s,i}=0,1$ and $l_{s,i}=0,1$. $\omega_{s,i}$ and $\mathbf{k}_{s,i\parallel}$ are frequency and transverse wave vector (in plane component of the wave vector of the plane wave mode) of the signal/idler photons, respectively. $j_{s,i}=0$ ($j_{s,i}=1$) presents the modes propagating upwardly (downwardly) respect to the $x$-$y$ plane. Here and henceforth, the "$\parallel$" in the subscript means the component is in $x$-$y$ plane. The electric fields of these modes are all linearly polarized, $l_{s,i}$ indicate two polarization directions. The modes with polarizations parallel with the $x$-$y$ plane are indicated by $l_{s,i}=0$, and $l_{s,i}=1$ indicates that the modes polarized along the plane spanned by the propagating direction and $z$-axis. In the following discussion, the two polarization directions are represented as "TE" and "TM", respectively. $\mathbf{k}_{s,i\parallel}$ can be expressed by the spatial frequency $k_{s,i\parallel}=|\mathbf{k}_{s,i\parallel}|$ and angle $\varphi_{ks,i\parallel}$, which are shown in Fig. 1 (b). The notation $|*|$ means the magnitude of a vector or the modulus of a complex expression. The integral variable $\mathbf{S}=[\mathbf{S}_s,\mathbf{S}_i]$ in Eq. (1) is a six-dimensional vector. $\psi(\mathbf{S},j_s,l_s,j_i,l_i)$ is the probability amplitude function for the signal and idler photons.

To simplify the calculation, we assume that the pump light is a series of Gaussian pulses with central frequency of $\omega_{p0}$ and temporal duration of $\tau_p$. Its spatial distribution is Gaussian, and its waist is on the ultrathin film with a radius of $r_p$. The peak power of the pump pulses is $P_p$. We also define two angular frequency detunings as $\Omega_{s,i}=\omega_{s,i}-\omega_{s,i0}$. Here, $\omega_{s,i0}$ is the central angular frequency of the signal or idler fields, which determined by the optical filters for the signal/idler photon collections and satisfied $\omega_{s0}+\omega_{i0}=2\omega_{p0}$. In this case $\psi(\mathbf{S},j_s,l_s,j_i,l_i)$ has an explicit expression (See the detailed derivation in Appendix A)

$$\psi(\Omega_s,\Omega_i,\mathbf{k}_{s\parallel},\mathbf{k}_{i\parallel},j_s,l_s,j_i,l_i)=\beta(\gamma P_p d)(\pi r_p^2)(\sqrt{2}\pi\tau_p)$$
$$\times G_\Omega(\Omega_s,\Omega_i)G_k(\mathbf{k}_{s\parallel},\mathbf{k}_{i\parallel})\Phi(\varphi_{ks},\varphi_{ki},l_s,l_i)$$
$$\times\Theta(\Omega_s,k_{s\parallel},j_s,l_s)\Theta(\Omega_i,k_{i\parallel},j_i,l_i)U(\Omega_s,k_{s\parallel})U(\Omega_i,k_{i\parallel}),$$
(2)

where $\gamma$ is the third order nonlinear coefficient proportional to $\chi^{(3)}_{xxxx}$, but inverse proportional to $r_p^2$. $\beta$ is a constant, and the explicit expressions of $\beta$ and $\gamma$ could be found in Appendix A.

The terms $G_\Omega(\Omega_s,\Omega_i)$ and $G_k(\mathbf{k}_{s\parallel},\mathbf{k}_{i\parallel})$ in Eq. (2) are

$$G_\Omega(\Omega_s,\Omega_i)=e^{-\tau_p^2(\Omega_s+\Omega_i)^2/4}, \quad (3.1)$$

$$G_{\mathbf{k}_{s\parallel}}(\mathbf{k}_{s\parallel},\mathbf{k}_{i\parallel})=e^{-r_p^2|\mathbf{k}_{s\parallel}+\mathbf{k}_{i\parallel}|^2/4}, \quad (3.2)$$

and they originate from the energy and transverse momentum conservations in the SpFWM, respectively.

The terms $\Phi(\varphi_{ks},\varphi_{ki},l_s,l_i)$ and $\Theta(\Omega_{s,i},k_{s,i\parallel},j_{s,i},l_{s,i})$ have forms of

$$\Phi(\varphi_{ks},\varphi_{ki},l_s,l_i)=$$
$$\sum_{m=0,1}(r_\chi)^m\cos(\varphi_{ks}+\frac{l_s-m}{2}\pi)\cos(\varphi_{ki}+\frac{l_i-m}{2}\pi), \quad (4.1)$$

$$\Theta(\Omega_{s,i},k_{s,i\parallel},j_{s,i},l_{s,i})=\cos^{l_{s,i}}[\theta_{s,i}(\Omega_{s,i},k_{s,i\parallel},j_{s,i})], \quad (4.2)$$

where $\theta_{s,i}(\Omega_{s,i},k_{s,i\parallel},j_{s,i})=(-1)^{j_{s,i}}\arcsin[k_{s,i\parallel}/k_{s,i}(\Omega_{s,i})]$ is the angle between the wave vector of a plane wave mode and the $z$-axis, in which $k_{s,i}(\Omega_{s,i})$ is its angular wavenumber at $\omega_{s,i}$. In all the calculations in this paper, a linear dispersion relationship $k_{s,i}=(\omega_{s,i0}+\Omega_{s,i})n_{i,s}/c$ is utilized, where $n_{s,i}$ is the refractive index of the ultrathin film. The parameter $r_\chi=\chi^{(3)}_{yyxx}/\chi^{(3)}_{xxxx}$ indicates the anisotropy of the nonlinear susceptibility.

The terms $U(\Omega_{s,i},k_{s,i\parallel})=u[k_{s,i}(\Omega_{s,i})/n_{s,i}-k_{s,i\parallel}]$ in Eq. (2) are unit step functions. It can be seen that if the value of in-plane wave vector component of a mode ($k_{s,i\parallel}$) is larger than $k_{s,i}(\Omega_{s,i})/n_{s,i}$, i.e. $k_{s,i\parallel}>k_{s,i}(\Omega_{s,i})/n_{s,i}$, the value of out-plane wave vector component of the mode ($k_{s,iz}$) would be imaginary, i.e., the square of $k_{s,iz}$ would be negative. It means that this mode is evanescent along $z$. In contrast, a mode with $k_{s,i\|} < k_{s,i}(\Omega_{s,i})/n_{s,i}$ has real $k_{s,iz}$ and it can propagate to the far field region. Hence, by introducing these unit step functions, the contributions of the evanescent modes on the expression of the biphoton state are removed, since the values of these functions are 0 for the evanescent modes and 1 for the propagating modes, respectively. It is reasonable since it can be expected that the photon pairs of the biphoton state would be measured at far field regions.

The effect of these unit step functions will be shown in following theoretical analyses about the spatial Schmidt number of this biphoton state. In these analyses, integrals of the state over the wave-vector space are required. Previous theoretical investigations on SPDC in bulk crystals has shown that the *sinc*-type phase mismatching factor in the biphoton states naturally plays as a filtering window in wave-vector space to avoid the divergence of these integrals. If the crystal is very thin and the *sinc*-type phase mismatching factor can be neglected in the biphton state, these integrals will be divergence due to lack of filtering windows in wave vector space [29]. It is obvious that similar problem also exists in the biphoton state generated by SpFWM in ultrathin materials, since there is also no *sinc*-type factor in Eq. (2). In this paper, we overcome this problem by introducing the unit step functions $U(\Omega_{s,i}, k_{s,i\|}) = u[k_{s,i}(\Omega_{s,i})/n_{s,i} - k_{s,i\|}]$ as the filtering windows in wave vector space, under the assumption that the generated photon pairs are measured at far field regions.

For a specific signal (idler) mode with transverse wave vector $\mathbf{k}_{s\|}$ ($\mathbf{k}_{i\|}$), the joint spectral intensity [30] of the biphoton state shown in Eq. (2) can be explicitly expressed as $F(\Omega_s, \Omega_i) = e^{-\tau_p^2(\Omega_s+\Omega_i)^2/2}$, which extends homogeneously along the axis of $\Omega_s = -\Omega_i$ until that $k_s(\Omega_s) \geq k_{s\|}$ and $k_i(\Omega_i) \geq k_{i\|}$ are not satisfied. It is clearly different from the $F(\Omega_s, \Omega_i)$ of the biphoton states generated by SPDC in thick nonlinear media, in which the phase mismatching term makes $F(\Omega_s, \Omega_i)$ decay along the axis of $\Omega_s = -\Omega_i$ [30]. This difference means that the frequency bandwidth of the biphoton state generated by SpFWM in an ultrathin film could be much larger than that generated by SPDC in thick nonlinear crystals.

### III. CORRELATION FUNCTION OF THE BIPHOTON STATE

The spatiotemporal structure of two photon correlation could be demonstrated by coincidence measurement of signal and idler photons under different time delay and spatial shift [16], which is in proportion to the modulus of correlation function of the biphoton state [15]. When the collection and detection processes of signal and idler photons are taken into consideration, the correlation function of the biphoton state described in Eq. (1) can be calculated by

$$C(t_s, t_i, \mathbf{r}_s, \mathbf{r}_i) = \langle 0|\langle 0|\hat{a}_s(t_s, \mathbf{r}_s)\hat{a}_i(t_i, \mathbf{r}_i)|\psi\rangle$$

$$= (2\pi)^{-3} \sum_{l_s=0,1}\sum_{l_i=0,1} \int d^6\mathbf{S}' \psi(\mathbf{S}, 0, l_s, 0, l_i) F_\Omega(\Omega_s) F_\Omega(\Omega_i)$$

$$\times F_k(k_{s\|}) F_k(k_{i\|}) e^{-i(\Omega_s t_s + \Omega_i t_i)} e^{i(\mathbf{k}_{s\|}\mathbf{r}_s + \mathbf{k}_{i\|}\mathbf{r}_i)}, \quad (5)$$

where $\hat{a}_{s,i}(t_{s,i}, \mathbf{r}_{s,i})$ (see the explicit expressions of them in Appendix B) is the annihilation operators of signal/idler photons at time $t_{s,i}$ and positions $\mathbf{r}_{s,i} = r_{s,i}[\cos\varphi_{s,i}, \sin\varphi_{s,i}]$. In the collection configuration for deriving Eq. (5), only the photons propagating upwardly with respect to x-y plane are collected, which results in the term $\psi(\mathbf{S}, j_s=0, l_s, j_i=0, l_i)$ inside the integral in Eq. (5). The terms $F_\Omega(\Omega_{s,i})$ and $F_k(k_{s,i\|})$ are introduced to represent the filtering processes on the frequency and spatial frequency in the collection and detection. In the calculation, they have forms of

$$F_\Omega(\Omega_{s,i}) = e^{-(\Omega_{s,i}^2)/2\Omega_c^2}, \quad (6.1)$$

$$F_k(k_{s,i\|}) = \begin{cases} 1 & \Upsilon_c k_{s,i\|} < 1/2 \\ 0 & \text{others} \end{cases}. \quad (6.2)$$

In Eq. (6.1), the signal and idler channels have the same frequency filtering bandwidth $\Omega_c$. The spatial filtering function shown in Eq. (6.2) is rectangular, centers at $k_{s,i\|}=0$ and gets cutoff at $k_{s,i\|}=1/2\Upsilon_c$. This function describes a filtering process in far field. It corresponds to a near field filtering function with variable of $r$, which is the radical coordinate in the x-y plane. The latter function is *sinc*-shaped, and the width of its main lobe is $4\pi\Upsilon_c$.

When $t_s$ and $\mathbf{r}_s$ are set as the origins of temporal and spatial coordinate, an explicit expression of Eq. (5) can be obtained as

$$C(\Delta t, \Delta \mathbf{r})$$

$$= C_c \sum_{l_s=0,1}\sum_{l_i=0,1} \int d\Omega_s \int d\Omega_i G_\Omega(\Omega_s, \Omega_i) F_\Omega(\Omega_s) F_\Omega(\Omega_i) e^{-i\Omega_i\Delta t}$$

$$\times \int k_{s\|} dk_{s\|} \int k_{i\|} dk_{i\|} g_k(k_{s\|}, k_{i\|}) I_1(\frac{r_p^2 k_{s\|} k_{i\|}}{2}) J(k_{i\|}, \Delta r, \varphi_i)$$

$$\times \Theta(\Omega_s, k_{s\|}, 0, l_s)\Theta(\Omega_i, k_{i\|}, 0, l_i) U(\Omega_s, k_{s\|}) U(\Omega_i, k_{i\|}), \quad (7)$$

where $\Delta t$ is the time delay and $\Delta \mathbf{r} = \Delta r e^{i\varphi_i}$ is the spatial shift between the idler and signal photons. In Eq. (7), $g_k(k_{s\|}, k_{i\|}) = e^{-r_p^2(k_{s\|}^2+k_{i\|}^2)/4}$ and $I_1(*)$ is the modified 1st-order Bessel function. The explicit expression of the constant $C_c$ can be found in Appendix B. The expression of $J(k_{i\|}, r_i, \varphi_i)$ is

$$J(k_{i\|}, \Delta r, \varphi_i) = -\frac{1}{2}(1+r_\chi)J_0(k_{i\|}\Delta r)\cos[\frac{\pi}{2}(m_s-m_i)]$$

$$+\frac{1}{2}(1-r_\chi)J_2(k_{i\|}\Delta r)\cos[\frac{\pi}{2}(m_s+m_i)+2\varphi_i], \quad (8)$$

where $J_0(*)$ and $J_2(*)$ are the 0th and 2nd-order Bessel functions, respectively. It is obvious that the correlation function in Eq. (7) has four terms, which describe the spatial and temporal correlations between the photon pair with polarizations of 'TE/TE', 'TE/TM', 'TM/TE' and 'TM/TM', respectively. In the following analysis, the four terms will be named as $C_{l_s l_i}(\Delta t, \Delta \mathbf{r})$ with $l_{s,i}=0,1$, and therefore $C(\Delta t, \Delta \mathbf{r}) = \sum_{l_s=0,1}\sum_{l_i=0,1} C_{l_s l_i}(\Delta t, \Delta \mathbf{r})$. All the four terms have the form of cascaded two dimensional Fourier transformation and two dimensional Hankel transformation.

If we consider a special case that the pump is treated as a monochromatic plane wave. A simple form of $C_{l_s l_i}(\Delta t, \Delta \mathbf{r})$ is obtained as

$$C_{l_s l_i}(\Delta t, \Delta \mathbf{r}) = C_c \int d\Omega_s F_\Omega^2(\Omega_s) e^{i\Omega_s \Delta t} \int k_{s\|} dk_{s\|}$$
$$\times F_k^2(k_{s\|}) J(k_{s\|}, \Delta r, \varphi_i) \Theta(\Omega_s, k_{s\|}, 0, l_s)$$
$$\times \Theta(\Omega_s, k_{s\|}, 0, l_i) U(\Omega_s, k_{s\|}). \quad (9)$$

This expression is the cascade of one dimensional Fourier transformation and one dimensional Hankel transformation. If $(\omega_{s0,i0} - \Omega_c) n_s / c \gg 1/\Upsilon_c$, i.e., the filtering bandwidths on the frequency and spatial frequency are not too large, $\Theta(\Omega_s, k_{s\|}, 0, l_s)$, $\Theta(\Omega_s, k_{s\|}, 0, l_i)$, and $U(\Omega_s, k_{s\|})$ are always 1. In this case, Eq. (9) can be further simplified to

$$C_{l_s l_i}(\Delta t, \Delta \mathbf{r}) = C_c \int d\Omega_s F_\Omega^2(\Omega_s) e^{i\Omega_s \Delta t}$$
$$\times \int k_{s\|} dk_{s\|} F_k^2(k_{s\|}) J(k_{s\|}, \Delta r, \varphi_i). \quad (10)$$

This equation is similar with the correlation function of the biphoton state generated in SPDC in bulk nonlinear crystal [31] pumped by plane wave, except two differences. First, the kernel function in the Hankel transformation is different. Second, Eq. (10) can be factored to two parts only depending on $\Delta t$ and $\Delta \mathbf{r}$, respectively, however, the correlation function in Ref. [31] does not have this property. Since the $\Delta t$-dependent part is the same for all the four $C_{l_s l_i}(\Delta t, \Delta \mathbf{r})$, the correlation function $C(\Delta t, \Delta \mathbf{r})$ of the biphoton state generated by SpFWM in ultrathin film can also be factored to temporal and spatial correlation functions. It means that the space and time in the biphoton state is factorable. Although this conclusion is obtained under the condition that $(\omega_{s0,i0} - \Omega_c) n_s / c \gg 1/\Upsilon_c$, in Fig.2 we will show that it is valid even when this condition is not satisfied well.

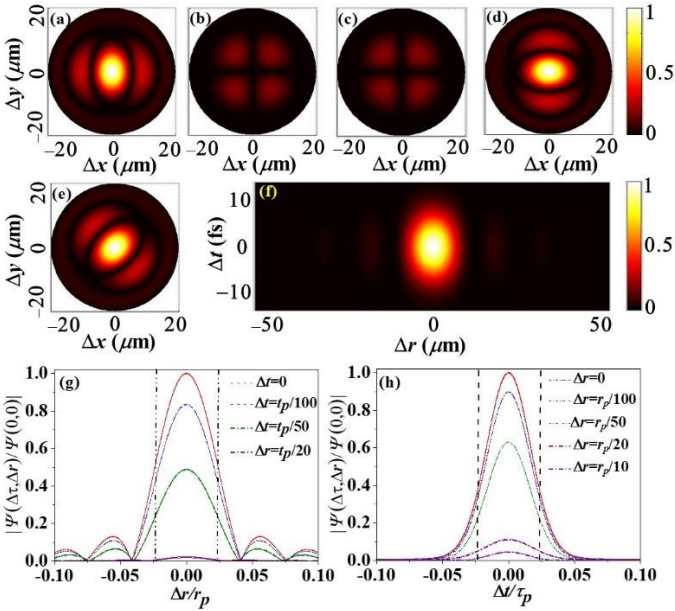

Fig. 2 The numerical calculations of the correlation function. (a)-(d) are $|C_{00}(0, \Delta \mathbf{r})|/|C_{00}(0,0)|$, $|C_{01}(0, \Delta \mathbf{r})|/|C_{00}(0,0)|$, $|C_{10}(0, \Delta \mathbf{r})|/|C_{00}(0,0)|$ and $|C_{11}(0, \Delta \mathbf{r})|/|C_{00}(0,0)|$ versus the spatial shifts $\Delta x = |\Delta \mathbf{r}|\cos\varphi_i$ and $\Delta y = |\Delta \mathbf{r}|\sin\varphi_i$. The vertical coordinates in (a)-(d) are the same, and the colorbar next to (d) is for all of (a)-(d). (e) The result of $|C(0, \Delta \mathbf{r})|$. (f) $|C(\Delta t, \Delta \mathbf{r})|/|C(0,0)|$ versus the $\Delta r$ and $\Delta t$, when $\varphi_i$ is set at $\pi/4$. The color bar next to (f) is for both of (e) and (f). The calculations about (a-f) are based on Eq. (9). (g) $|C(\Delta t, \Delta r)|/|C(0,0)|$ versus $\Delta r$ under different $\Delta t$. (h) $|C(\Delta t, \Delta r)|/|C(0,0)|$ versus $\Delta t$ under different $\Delta r$. The calculations for (g) and (h) are based on Eq. (7). In (g) and (h), $\tau_p = 1$ps, $r_p = 10\mu m$, respectively. In all of the calculations, the central wavelength of the pump and idler fields are set as $\lambda_{p0} = 1530$nm and $\lambda_{i0} = 1310$nm, respectively. Other parameters are $n_{i,s} = 2.6$, $r_\chi = 1/3$, $\Omega_c = 2\pi \times 50$THz and $\Upsilon_c = 1\mu m$.

According to Eq. (9), we calculated all the four $C_{l_s l_i}(0, \Delta \mathbf{r})$ under $\Omega_c = 2\pi \times 50$THz, $\Upsilon_c = 1\mu m$ and $n_{i,s} = 2.6$. The anisotropy of the nonlinear film is set by $r_\chi = 1/3$. In this condition, $(\omega_{s0,i0} - \Omega_c) n_s / c \gg 1/\Upsilon_c$ is not satisfied. The normalized results are shown in Figs. 2 (a)-(d). It is obvious that all the four terms are azimuthally anisotropic since the anisotropy of the nonlinear susceptibility is considered in our model. It is different from the isotropic correlation function in Ref. [31]. Moreover, $|C_{00}(0, \Delta r)|$ and $|C_{11}(0, \Delta r)|$ are much larger than $|C_{10}(0, \Delta r)|$ and $|C_{01}(0, \Delta r)|$, hence, the former two are dominant in the calculation of $|C(0, \Delta r)|$. It is obvious that the orientations of $|C_{00}(0, \Delta r)|$ and $|C_{11}(0, \Delta r)|$ is along $\varphi_i = 0$ and $\varphi_i = \pi/2$, respectively. It makes $|C(0, \Delta \mathbf{r})|$ nearly orient along $\varphi_i = \pi/4$, which is calculated according to Eq. (9) and shown in the Fig. 2 (e). In Fig. 2 (f), $|C(\Delta t, \Delta r)|/|C(0,0)|$ versus $\Delta t$ and $\Delta r$ are shown with $\varphi_i$ at $\pi/4$ and $\Delta \mathbf{r} = \Delta r e^{i\pi/4}$. When $\Delta t$ ($\Delta r$) is fixed at arbitrary position, $|C(\Delta t, \Delta r)|$ always shows Airy disk section (Gaussian) profile with variable of $\Delta r$ ($\Delta t$), which corresponds to the profiles of the Hankel (Fourier) transformation of the filtering function in Eq. (6.1) (Eq. (6.2)). These results show that biphoton correlation of the state in Eq. (1) is factorable in temporal and spatial domains even under larger filtering bandwidths in frequency and spatial frequency. This property is much different from the biphoton state generated by SPDC in bulk nonlinear crystals. Their spatiotemporal structure of the biphoton correlation has nonfactorable *X*-geometry [31]. Such nonfactorability is determined by the requirement of phase match of SPDC in bulk crystals. On the other hand, for the biphoton state generated by SpFWM in ultrathin films, the factorability in the biphoton correlation revealed in Fig. 2 is due to the relaxation of the requirement of longitudinal phase match.

In the calculations for Fig. 2 (f), the pump beam is approximated by monochromatic plane wave. By numerically calculating the Eq. (7), we can study whether the biphoton correlation is factorable under Gaussian pump. The Monte Carlo method is employed to numerically compute the integrals in Eq. (7), when $\tau_p = 1$ps, $r_p = 10\mu m$ and other parameters are the same to those used in the calculation for Fig. 2 (f). The calculated results are shown in Figs. 2 (g) and (h). In Fig. 2 (g), the calculated normalized correlation function $|C(\Delta t, \Delta r)/C(0,0)|$ versus $\Delta r$ is shown under different $\Delta t$. It can be seen that varying $\Delta t$ does not change the Airy disk section profile of the calculated curves and its full width at half maxima (FWHM), which is the correlation length of the biphoton state. It only leads to the variation of magnitude of the curves. Figure 2 (h) shows the calculated $|C(\Delta t, \Delta r)/C(0,0)|$ versus $\Delta t$, when $\Delta r$ is set at different values. It can be seen that all the curves under different $\Delta r$ have Gaussian profiles. Their widths indicate the correlation time of the biphoton state, which are almost unchanged under different $\Delta r$. Only the magnitudes of these curves decreases with increasing $\Delta r$. The two figures indicate that the biphoton correlation is still factorable in temporal and spatial domains even that pump light with Gaussian field distribution is considered.

## IV. SPATIOTEMPORAL SEPARABILITY OF THE BIPHOTON STATE

As shown in Fig. 2, the correlation function of the biphoton generated in ultrathin film has factorable spatiotemporal structure. Since the correlation function is calculated based on the expression of the biphoton state, the temporal and spatial degrees of freedom in the biphoton state itself would be separable. The separability of a quantum state in two degrees of freedom can be evaluated by the purity of quantum state when one of the degrees of freedom is traced out [21]. To explore the spatio-temporal separatiblity of the biphoton state shown in Eq. (1), we calculate the frequency purity $\mathbf{Tr}(\rho^2_{\omega_s,\omega_i})$ by tracing out the both of the spatial and polarization degrees of freedom. The result is (See the detailed derivation in Appendix C)

$$\mathbf{Tr}(\rho^2_{\omega_s,\omega_i}) = \frac{B}{N^2}, \quad (11)$$

where

$$N = \frac{1}{4}\sum_{l_s=0,1}\sum_{l_i=0,1}\int_{-\infty}^{\infty}d\Omega_s\int_{-\infty}^{\infty}d\Omega_i\int k_{s\parallel}dk_{s\parallel}\int k_{i\parallel}dk_{i\parallel}$$
$$\times G_\Omega^2(\Omega_s,\Omega_i)g_k^2(k_{s\parallel},k_{i\parallel})F_\Omega^2(\Omega_s)F_\Omega^2(\Omega_i)F_k^2(k_{s\parallel})F_k^2(k_{i\parallel})$$
$$\times I(k_{s\parallel},k_{i\parallel},l_s,l_i)\Theta^2(\Omega_s,k_{s\parallel},0,l_s)\Theta^2(\Omega_i,k_{i\parallel},0,l_i)$$
$$\times U(\Omega_s,k_{s\parallel})U(\Omega_i,k_{i\parallel}), \quad (12)$$

and

$$B = \sum_{l_{s1}=0,1}\sum_{l_{i1}=0,1}\sum_{l_{s2}=0,1}\sum_{l_{i2}=0,1}\int_{-\infty}^{\infty}d\Omega_{s1}\int_{-\infty}^{\infty}d\Omega_{i1}\int_{-\infty}^{\infty}d\Omega_{s2}\int_{-\infty}^{\infty}d\Omega_{i2}$$
$$\times \int k_{s1\parallel}dk_{s1\parallel}\int k_{s2\parallel}dk_{s2\parallel}\int k_{i1\parallel}dk_{i1\parallel}\int k_{i2\parallel}dk_{i2\parallel}$$
$$\times [\prod_{m=1,2}G_\Omega^2(\Omega_{sm},\Omega_{im})g_k^2(k_{sm\parallel},k_{im\parallel})F_\Omega^2(\Omega_{sm})F_\Omega^2(\Omega_{im})$$
$$\times F_k^2(k_{sm\parallel})F_k^2(k_{im\parallel})I(k_{sm\parallel},k_{im\parallel},l_{sm},l_{im})]$$
$$\times [\prod_{\alpha=1,2}\prod_{\beta=1,2}\Theta^2(\Omega_{s\alpha},k_{s\beta\parallel},0,l_{s\beta})$$
$$\times \Theta^2(\Omega_{i\alpha},k_{i\beta\parallel},0,l_{i\beta})U(\Omega_{s\alpha},k_{s\beta\parallel})U(\Omega_{i\alpha},k_{i\beta\parallel})], \quad (13)$$

In Eqs. (12) and (13), the term $I(k_{s\parallel},k_{i\parallel},l_s,l_i)$ is defined as

$$I(k_{s\parallel},k_{i\parallel},l_s,l_i) = (1+r_\chi^2)I_0(r_p^2k_{s\parallel}k_{i\parallel})$$
$$+(1+r_\chi)^2\cos[(l_s-l_i)\pi]I_2(r_p^2k_{s\parallel}k_{i\parallel}). \quad (14)$$

Compared with the denominator in Eq. (11), the numerator has eight additional terms, four cosine and four unit step functions in the integral. All the eight terms are smaller than 1, which guarantees $\mathbf{Tr}(\rho^2_{\omega_s,\omega_i})\leq 1$ according to Eq. (11). When $\Omega_c$ is small and $\Upsilon_c$ is relatively large, i.e., the detection process is narrowband in both of the frequencies and spatial frequencies, all of the eight additional terms are nearly unit. In such case, the denominator and numerator in Eq. (10) is equal, leading to $\mathbf{Tr}(\rho^2_{\omega_s,\omega_i})=1$. It shows that the biphoton state are separable in its temporal and spatial degrees of freedoms. This conclusion is consistent with the limit case shown in Ref. [21], where the two degrees of freedoms are separable in the biphoton state generated by SPDC in bulk crystal when the filtering bandwidth of the frequency is nearly 0 or that of the spatial frequency is very small.

When the bandwidths of the frequency and spatial frequency filtering processes increases, Eq. (10) cannot be simplified in the way mentioned above. We calculate the evolution of the frequency purity with increasing $\Upsilon_c$ and under different $\Omega_c$ numerically according to Eqs. (10), (11), and (12). The results are shown in Fig. 3. The multi-dimensional integrals in Eq. (11) and (12) are implemented by the Mente Carlo methods. In the calculations, the rectangular filtering function in Eqs. (6.2) and (11) is replaced by $F_k(k_{s,i\parallel}) = e^{-\Upsilon_c^2 k_{s,i\parallel}^2/2}$, to make the comparison between our results and the results of similar calculation for biphoton state generated by SPDC in bulk crystal [21] reasonable, in which the filtering function of spatial frequency is Gaussian.

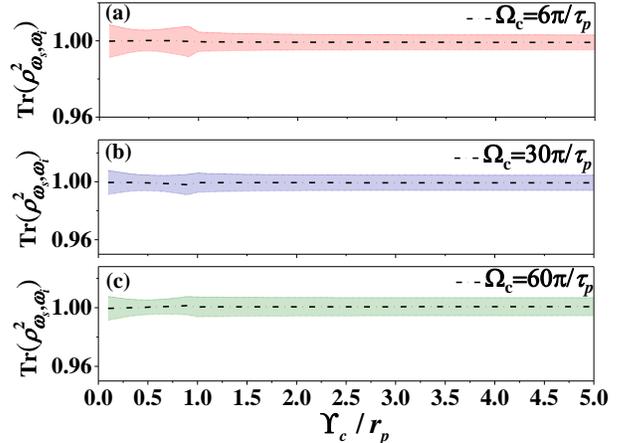

Fig. 3 The numerical calculations of the frequency purity $\mathbf{Tr}(\rho^2_{\omega_s,\omega_i})$. In (a), (b) and (c), the purity versus $\Upsilon_c$ is calculated when $\Omega_c$ are set at $6\pi/\tau_p$, $30\pi/\tau_p$, and $60\pi/\tau_p$, respectively. All the other parameters in the calculation for (a), (b) and (c) are the same to those used in Fig. (2). In each figure, the shadow band shows the results of 50 times of numerical computations involving Monte Carlo integrals for each case, the dashed dot line is their average.

Figure 3 (a) is the result under different $\Upsilon_c$ when the frequency filtering bandwidth is set 3THz. The shadow band shows the results of 50 times of numerical computations involving Monte Carlo integrals for each case, and the dashed dot curve is their average. It can be seen that the calculated purity $\text{Tr}(\rho^2_{\mathbf{k}_{s\parallel},\mathbf{k}_{i\parallel}})$ is always close to 1 under varying $\Upsilon_c$. Figure 3 (b) and (c) show the results when the frequency filtering bandwidth increases to 30THz and 60 THz, respectively. It can be seen that the purity is still close to 1 with broader $\Omega_c$. These results indicate that the temporal and spatial degrees of freedoms of the biphoton state studied here are separable, even under such large temporal and spatial filtering bandwidths. As a comparison, previous works of the analysis on biphoton states generated by SPDC in bulk crystals [11] shows that, for the biphoton state generated in a crystal with a length of 1mm, the state purity would drop to near 0.2 when $\Omega_c/2\pi$=4.7THz and $\Upsilon_c = 100\mu m$. Hence, it is concluded that the biphoton states generated by SpFWM in ultrathin film are spatiotemporally separatable even under large photon collection/detection frequency and spatial frequency bandwidths, thanks to the relaxation of the requirement on longitudinal phase match. Since the separability between different degrees of freedom is important for the demonstration of hyper-entanglement, therefore ultrathin film is a type of promising material for realizing biphoton hyper-entanglement [21].

## V. HIGH DIMENSIONAL TRANSVERSE ENTANGLEMENT IN BIPHOTON STATE

Above analysis shows that the requirement of longitude phase match relaxes in SpFWM in ultrathin films. On the other hand, the transverse phase match in this process is still required. It leads to the property of high dimensional transverse entanglement, which can be indicated by the spatial Schmidt number. Larger spatial Schmidt number means higher entanglement dimension [29]. The spatial Schmidt number can be obtained by Schmidt decomposition of the biphoton state under the bases of plane wave modes [29, 32]. Here, we calculated the spatial Schmidt number to show the high dimensional transverse entanglement in the biphoton state generated by SpFWM in an ultrathin film.

Firstly, the spatial Schmidt number is obtained by $K_F = 1/\text{Tr}[\rho^2(\mathbf{k}_{s\parallel},\mathbf{k}_{i\parallel})]$, in which $\rho(\mathbf{k}_{s\parallel},\mathbf{k}_{i\parallel})$ is the density matrix of the biphoton state in the subspace spanned by $\mathbf{k}_{s\parallel}$ and $\mathbf{k}_{i\parallel}$ [11]. Under the assumption that only the upwardly propagating photons with specific signal and idler frequencies are collected, the explicit expression of $K_F$ can be deduced (see the detailed derivation in Appendix D) from the expression of the biphoton state in Eq. (2), and it has form of

$$K_F = \frac{N_K^2}{T}, \quad (15)$$

where

$$N_K = \sum_{l_s=0,1}\sum_{l_i=0,1}\int d\mathbf{k}_{s\parallel}\int d\mathbf{k}_{i\parallel}\,|\psi(\mathbf{k}_{s\parallel},\mathbf{k}_{i\parallel},l_s,l_i)|^2, \quad (16)$$

and

$$T = \sum_{l_{s1}=0,1}\sum_{l_{i1}=0,1}\sum_{l_{s2}=0,1}\sum_{l_{i2}=0,1}\int d\mathbf{k}_{s1\parallel}\int d\mathbf{k}_{s2\parallel}\int d\mathbf{k}_{i1\parallel}\int d\mathbf{k}_{i2\parallel}$$
$$\times \psi(\mathbf{k}_{s1\parallel},\mathbf{k}_{i1\parallel},l_{s1},l_{i1})\psi^*(\mathbf{k}_{s2\parallel},\mathbf{k}_{i1\parallel},l_{s1},l_{i1})$$
$$\times \psi(\mathbf{k}_{s2\parallel},\mathbf{k}_{i2\parallel},l_{s2},l_{i2})\psi^*(\mathbf{k}_{s4\parallel},\mathbf{k}_{i2\parallel},l_{s2},l_{i2}). \quad (17)$$

The $\psi(\mathbf{k}_{s\parallel},\mathbf{k}_{i\parallel},l_s,l_i)$ in Eqs. (15) and (16) is the $\psi(\Omega_s,\Omega_i,\mathbf{k}_{s\parallel},\mathbf{k}_{i\parallel},j_s,l_s,j_i,l_i)$ in Eq. (3) with fixed $\Omega_s$, $\Omega_i$ and $j_{s,i}=1$. Although the implementation of the integrals about $\varphi_{ks}$ and $\varphi_{ki}$ in Eqs. (15) and (16) can give expressions (the Eqs. (D9-D18) in Appendix D) appropriate for calculating $K_F$ numerically, they are too complex to give intuitionistic physical picture. Hence, we take two assumptions. First, only the photon pairs with "TE/TE" polarizations are collected. Second, $r_p$ is relatively large so that the term $G_{\mathbf{k}_{s\parallel}}(\mathbf{k}_{s\parallel},\mathbf{k}_{i\parallel})$ in $\psi(\mathbf{k}_{s\parallel},\mathbf{k}_{i\parallel},l_s,l_i)$ can be approximated by $4\pi\delta(\mathbf{k}_{s\parallel}+\mathbf{k}_{i\parallel})/r_p^2$. Under these assumptions, we get an approximation of $K_F$ as (See detailed derivation in Appendix D)

$$K_F = \frac{1}{8}\frac{\int d\mathbf{k}_{s\parallel} U(\Omega_s,k_{s\parallel})U(\Omega_i,k_{s\parallel})}{\frac{2\pi}{r_p^2}}. \quad (18)$$

On the other hand, an approximated expression of the spatial Schmidt number of the biphoton state generated by SPDC in thick crystal can be obtained as (See detailed derivation in Appendix D)

$$K_S = \frac{3}{8}\frac{\int d\mathbf{k}_{s\parallel}\,\text{sinc}^2(\Delta k L)U(\Omega_s,k_{s\parallel})U(\Omega_i,k_{s\parallel})}{\frac{2\pi}{r_p^2}}, \quad (19)$$

where $\Delta k$ and $L$ is the longitudinal phase mismatch in the SPDC and the thickness of the second order nonlinear crystal. The comparison between Eqs. (18) and (19) shows clear physics. In Eq. (18), the numerator has contributions of all the modes propagating along the longitudinal direction, while in Eq. (19), the numerator only has contributions of the modes satisfying the longitudinal phase matching condition. The denominators in Eqs. (18) and (19) are same and define the correlation volume [32]. Since the numerator in Eq. (19) is much smaller than that in Eq. (18), $K_F$ is much larger than $K_S$ under the same $r_p$. Therefore, the dimension of the transverse entanglement in the biphoton state generated by SpFWM in ultrathin films could be much higher than that of the state generated by SPDC in bulk crystal, due to the disappearance of the longitudinal phase mismatch. This conclusion is based on the two assumptions leading to Eq. (18), but in Fig. 4 it will be further verified by numerical calculations based on the Eq. (D9-D18) in Appendix D without any approximation.

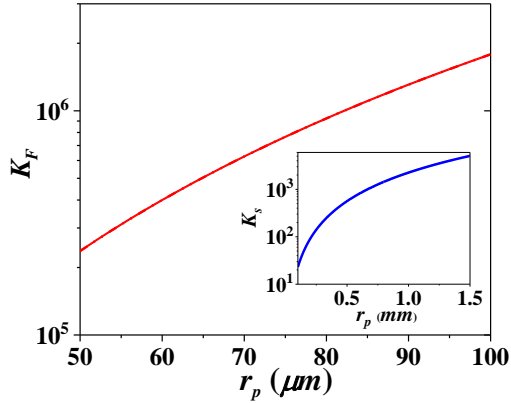

Fig. 4 The numerical calculations of the spatial Schmidt number $K_F$ of the biphoton state generated by SpFWM in ultrathin film. The inset shows $K_S$ for biphoton state generated by SPDC in a nonlinear crystal with a length of 1mm. In the calculations of SpFWM, $\lambda_p = 1530$nm, $\lambda_s = 1580$nm and $n_{i,s} = 2.6$. In the calculations of SPDC, $\lambda_p = 405$nm and $\lambda_s = \lambda_i = 810$nm, and $n_s = n_i = 1.7$.

According to Eq. (15-17), the spatial Schmidt number $K_F$ is calculated numerically and the results are shown in Fig. 4. It can be seen that $K_F$ increases with increasing $r_p$. According to Eq. (18), $K_F$ can be considered as the ratio between a certain volume and the correlation volume determined by $r_p$. Larger $r_p$ gives lower correlation volume and leads to higher $K_F$. The spatial Schmidt number $K_S$ versus $r_p$ for the biphoton state generated by SPDC in a nonlinear crystal with a length of $L$=1mm is also calculated according to the Eq. (23) in Ref. [11], and shown in Fig. 4 for comparison. The comparison between the two curves also show that even with smaller $r_p$, $K_F$ could be much higher than $K_S$.

## VI. DISCUSSION ON EXPERIMENTAL DEMONSTRATION

Coincidence photon counting measurement is the way to experimentally investigate the properties of biphoton states generated by SpFWM in ultrathin films. Since the nonlinear interaction length in an ultrathin film is very small, the key to demonstrate the biphoton state generation is whether sufficient coincidence photon counts can be recorded using proper materials and reasonable pump conditions. Several types of films, such as $Bi_2Se_3$ and graphene, have proven their high third-order nonlinearity [33, 34]. In the following discussions, we will estimate the coincidence photon count rates of the photon pairs generated by SpFWM in single/few-layered $Bi_2Se_3$ and graphene.

Considering that a pulsed pump light illuminates the ultrathin film in the vertical direction. It is a Gaussian beam with a Gaussian temporal pulse profile. Its waist locates on the film. The generated signal and idler photons are collected by a multi-mode collection system. Specifically, they are collected by lens with angular aperture of $\theta_c$, after they pass through corresponding frequency filters with unified angular frequency bandwidth of $\Omega_c$, and finally detected by two single photon detection devices with effective area of $\pi r_c^2$. If $\theta_c$ and $\Omega_c$ are not too large, the coincidence photon count rate $n_c$ per pulse can be estimated by the following equation (see the detailed derivation in Appendix E)

$$n_c \approx \alpha_c \eta_s \eta_i (\gamma P_p d)^2 \Omega_c \tau_p \sin^2\frac{\theta_c}{2}(\frac{r_c}{\lambda_{p0}})^2$$
$$\times [(\frac{1+r_\chi}{2})^2 + (\frac{1-r_\chi}{2})^2], \quad (20)$$

where $\alpha_c = n_{p0}^2 \pi^2 / 9 n_s n_i$, $\eta_s$ ($\eta_i$) and $\lambda_{p0}$ are the detection efficiency of the signal (idler) photons and the central wavelength of the pump light, respectively. $P_p$ and $\tau_p$ are the peak power and duration of the pump pulses, respectively.

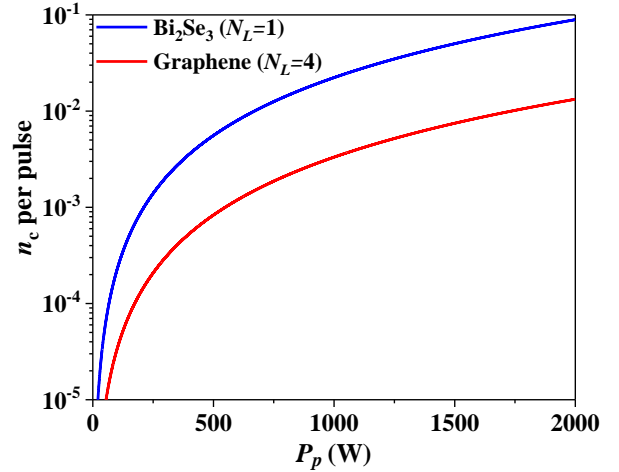

Fig. 5 Calculation results of signal-idler coincidence photon count rate $n_c$ per pulse versus $P_p$, in $Bi_2Se_3$ (layer number $N_L$=1) and graphene ($N_L$=4). The pulse width $\tau_p$ and spatial radius $r_p$ of the pump light are set at 200fs and 5$\mu$m, respectively. The pump and signal wavelengths, nonlinear susceptibilities or nonlinear refractive indices, refractive indices, absorption coefficients, as well as other parameters used in the calculation are available in Table F1 the Appendix F.

According to Eq. (20), the coincidence count rate $n_c$ per pulse is calculated for single-layered $Bi_2Se_3$ and four-layered graphene. Figure 5 shows the calculated $n_c$ per pulse as a function of the peak power $P_p$ of the pump pulses. It can be seen that $n_c$ increases significantly with raised $P_p$. For $Bi_2Se_3$ and graphene, the coincidence count rate per exceeds 0.01 per pulse when $P_p$ gets higher than 0.67kW and 1.73 kW. In the calculations about Fig. 5, $r_p$ and $\tau_p$ are set at 5$\mu$m and 200fs, respectively. The possible laser induced damage in materials also should be considered in experiment. In Ref. [34], $Bi_2Se_3$ shows no damage even when the peak intensity of the incident laser pulses reaches 10.4GW/cm$^2$, corresponding to $P_p$=8.2kW with $r_p$=5$\mu$m. Moreover, the damage threshold of graphene is around 14mJ/cm$^2$ [35], which corresponds to $P_p$=55kW, with $r_p$ and $\tau_p$ mentioned above. Therefore, $P_p$ used in estimating $n_c$ per pulse in Fig. 5 are far lower than the damage thresholds of the two materials. These calculations indicate that the biphoton state generated by SpFWM in ultrathin films could be demonstrated by proper material selection and reasonable pumping condition, with considerable coincidence count rates for applications on quantum optics and quantum information involving high-dimensional entanglement.

Under the pump power levels used in the calculation in Fig. 5, Kerr effect might become significant and its influence should be taken into consideration. Since the longitudinal phase mismatching term in the expression of the biphoton state disappears in ultra-thin materials, the change of the refractive index due to Kerr effect may influence the biphoton state only via the unit step functions presenting the truncation of the transverse wave vectors. However, in the numerical calculations of correlation functions in Fig. 2 and frequency purities in Fig. 3, as well as the evaluations of coincidence count in Fig. 5, the unit step function is always 1 within filtering bandwidths. Hence, the Kerr effect has little impact on the results shown in these three figures. Also, it can be calculated that the refractive indices of the signal (idler) fields in $Bi_2Se_3$ and graphene will increase to 4.8 (6.4) and 5.4 (5.4) due to Kerr effect when $P_p$ are set at 2kW. According to Eq. (18), the spatial Schmidt number $K_F$ will increase by 2.2 and 3.27, respectively. This indicate that the numerical calculation of $K_F$ in Fig. 4 would get higher in real experiment due to the probably existing Kerr effect.

In conclusion, we have theoretically investigated the biphoton state generated by SpFWM in ultrathin nonlinear films. The expression of the biphoton state is obtained by perturbation method under low parametric gain. Thanks to the ultrathin film structure, the requirement of longitudinal phase match is relaxed, supporting very broad frequency bandwidth of the biphoton state. On the other hand, the spatial frequency in the expression is truncated to keep only the longitudinal propagating modes, by which the divergence problem mentioned in Ref. [29] can be overcome. Then, the spatial-temporal structure of the correlation function of the state is analyzed, showing that the space and time in the biphoton correlation are factorable due to the relaxation of the longitude phase match. By analyzing the state purity when the space and polarization degrees of freedom are traced out, the time-space factorability of the biphoton state is also demonstrated numerically. We also calculated the spatial Schmidt numbers of the state, indicating that high dimensional transverse entanglement could be realized by this way. Finally, the feasibility of experimental demonstration of this biphoton state generation is evaluated numerically using some typical nonlinear ultrathin materials, showing its potential on applications involving hyper-entanglement and high dimensional entanglement.

## ACKNOWLEDGMENTS
This work was supported by the National Key R&D Program of China under Contract No. 2017YFA0303704 and 2018YFB2200400, National Natural Science Foundation of China under Contract Nos. 61575102, 91750206 and 61621064, Beijing National Science Foundation under Contract No. Z180012, Beijing Academy of Quantum Information Sciences under Contract No. Y18G26.

## APPENDIX A: DERIVATION OF THE EXPRESSION OF THE BIPHOTON STATE

In the interaction, the pump light is treated classically with Gaussian field distribution on the thin film. It is linearly polarized and the polarization is parallel to the $x$ axis. The electrical field intensity of the pump light can be expressed as:

$$\mathbf{E}_p(t,x,y,z) = \hat{\mathbf{x}} E_p^{(+)}(t,x,y,z) + \hat{\mathbf{x}} E_p^{(-)}(t,x,y,z)$$
$$= \hat{\mathbf{x}} E_0 e^{-\frac{t^2}{2\tau_p^2}} e^{-\frac{(x^2+y^2)}{2r_p^2}} e^{i(k_{pz}z-\omega_{p0}t)} + \hat{\mathbf{x}} E_0 e^{-\frac{t^2}{2\tau_p^2}} e^{-\frac{(x^2+y^2)}{2r_p^2}} e^{-i(k_{pz}z-\omega_{p0}t)}, \quad (A1)$$

where $\omega_{p0}$, $\tau_p$ and $r_p$ are the central angular frequency, temporal duration and the transverse radius of the pump beam. Here, $\hat{\mathbf{x}}$ is the unit vector along $x$ axis. After implementing the Fourier transformation for $t$, $x$ and $y$, we express the electrical filed in the domains of frequency and spatial frequencies as

$$\mathbf{E}_p(\omega_p, k_{px}, k_{py}, z) = \hat{\mathbf{x}} E_p^{(+)}(\omega_p, k_{px}, k_{py}, z) + \hat{\mathbf{x}} E_p^{(-)}(\omega_p, k_{px}, k_{py}, z)$$
$$= \hat{\mathbf{x}} E_0 \tau_p r_p^2 e^{-\frac{r_p^2(k_{px}^2+k_{py}^2)}{2}} [e^{-\frac{\tau_p^2(\omega_p-\omega_{p0})^2}{2}} e^{ik_{pz}z} + e^{-\frac{\tau_p^2(\omega_p+\omega_{p0})^2}{2}} e^{-ik_{pz}z}], \quad (A2)$$

where $\omega_p$, $k_{px}$ and $k_{py}$ are angular frequency, the $x$ and $y$ component of the transverse wave vector of the pump field, respectively.

The quantized signal and idler fields generated in the SpFWM have expressions [36, 37] of

$$\hat{\mathbf{E}}_s(t,x,y,z) = \hat{\mathbf{E}}_s^{(+)}(t,x,y,z) + \hat{\mathbf{E}}_s^{(-)}(t,x,y,z)$$
$$= (2\pi)^{-\frac{3}{2}} i \int d\omega_s \sqrt{\frac{\hbar\omega_s}{2c\varepsilon_0 n_s}} \int dk_{sx} \int dk_{sy} \sum_{j_s=0,1} \sum_{l_s=0,1} \hat{\mathbf{e}}_{s,l_s} \hat{a}_s(\omega_s, k_{sx}, k_{sy}, j_s, l_s) e^{i(k_{sx}x+k_{sy}y)} e^{i[(-1)^{j_s} k_{sz}z - \omega_s t]} u[k_s(\omega_s)/n_s - k_{s\parallel}]$$
$$- (2\pi)^{-\frac{3}{2}} i \int d\omega_s \sqrt{\frac{\hbar\omega_s}{2c\varepsilon_0 n_s}} \int dk_{sx} \int dk_{sy} \sum_{j_s=0,1} \sum_{l_s=0,1} \hat{\mathbf{e}}_{s,l_s} \hat{a}_s^\dagger(\omega_s, k_{sx}, k_{sy}, j_s, l_s) e^{-i(k_{sx}x+k_{sy}y)} e^{-i[(-1)^{j_s} k_{sz}z - \omega_s t]} u[k_s(\omega_s)/n_s - k_{s\parallel}], \quad (A3)$$

$$\hat{\mathbf{E}}_i(t,x,y,z) = \hat{\mathbf{E}}_i^{(+)}(t,x,y,z) + \hat{\mathbf{E}}_i^{(-)}(t,x,y,z)$$
$$= (2\pi)^{-\frac{3}{2}} i \int d\omega_i \sqrt{\frac{\hbar\omega_i}{2c\varepsilon_0 n_i}} \int dk_{ix} \int dk_{iy} \sum_{j_i=0,1} \sum_{l_i=0,1} \hat{\mathbf{e}}_{i,l_i} \hat{a}_i(\omega_i, k_{ix}, k_{iy}, j_i, l_i) e^{i(k_{ix}x+k_{iy}y)} e^{i[(-1)^{j_i} k_{iz}z - \omega_i t]} u[k_i(\omega_i)/n_i - k_{i\parallel}]$$
$$- (2\pi)^{-\frac{3}{2}} i \int d\omega_i \sqrt{\frac{\hbar\omega_i}{2c\varepsilon_0 n_i}} \int dk_{ix} \int dk_{iy} \sum_{j_i=0,1} \sum_{l_i=0,1} \hat{\mathbf{e}}_{i,l_i} \hat{a}_i^\dagger(\omega_i, k_{ix}, k_{iy}, j_i, l_i) e^{i(k_{ix}x+k_{iy}y)} e^{-i[(-1)^{j_i} k_{iz}z - \omega_i t]} u[k_i(\omega_i)/n_i - k_{i\parallel}], \quad (A4)$$

where $\omega_s$ ($\omega_i$), $k_{sx}$ ($k_{ix}$) and $k_{sy}$ ($k_{iy}$) are the angular frequency, the $x$-component and $y$-component of the transverse wave vectors of the signal (idler) field, respectively. Here, $n_s$ and $n_i$ are the refractive indices of the signal and idler fields, respectively. The meaning of $j_{s,i}$ and $l_{s,i}$ have been given in the paragraph below Eq. (1) in Sec. II. The four indices $\omega_{s,i}$, $k_{sx,ix}$, $k_{sy,iy}$ and $j_{s,i}$ can determine the wave vector of a mode uniquely. For instance, if a mode has indices $\omega_s$, $k_{sx}$ and $k_{sy}$, its wave vector can be represented as ($k_{sx}$, $k_{sy}$, $(-1)^j \sqrt{k_s^2 - k_{sx}^2 + k_{sy}^2}$) in the Cartesian coordinate for the upwardly and downwardly propagating modes, respectively, where $k_s$ is the total angular wavenumber and depend on $\omega_s$ via the dispersion relationship $k_s(\omega_s)$. In the summation in Eq. (A3) and (A4), $l_{s,i}=0$ and 1 presents the "TE" and "TM" polarizations, respectively. The vector $\hat{\mathbf{e}}_{s,0}$ and $\hat{\mathbf{e}}_{s,1}$ ($\hat{\mathbf{e}}_{i,0}$ and $\hat{\mathbf{e}}_{i,1}$) are the unit vectors along the "TE" and "TM" polarization direction for signal (idler) photons. The operators of $\hat{a}_s(\omega_s, k_{sx}, k_{sy}, j_s, l_s)$ ($\hat{a}_i(\omega_i, k_{ix}, k_{iy}, j_i, l_i)$) and $\hat{a}_s^\dagger(\omega_s, k_{sx}, k_{sy}, j_s, l_s)$ ($\hat{a}_i^\dagger(\omega_i, k_{ix}, k_{iy}, j_i, l_i)$) are the annihilation and creation operators of the signal (idler) fields, respectively. It is notable that the dimensions of $\langle \hat{a}_{s,j,l}^\dagger(\omega_s, k_{sx}, k_{sy}) \hat{a}_{s,j,l}(\omega_s, k_{sx}, k_{sy}) \rangle$ ($\langle \hat{a}_{i,j,l}^\dagger(\omega_i, k_{ix}, k_{iy}) \hat{a}_{i,j,l}(\omega_i, k_{ix}, k_{iy}) \rangle$) is the photon number within unit volume in the space spanned by $\omega_s$, $k_{sx}$ and $k_{sy}$ ($\omega_i$, $k_{ix}$ and $k_{iy}$). In Eq. (A3) and (A4), $u[k_s(\omega_s)/n_s - k_{s\parallel}]$ ($u[k_s(\omega_s)/n_s - k_{s\parallel}]$) is a unit step function, in which $k_{s\parallel} = \sqrt{k_{sx}^2 + k_{sy}^2}$ ($k_{i\parallel} = \sqrt{k_{ix}^2 + k_{iy}^2}$) is the total transverse angular wavenumber of the signal (idler) filed. By the two unit step functions, only the propagating modes are kept in the calculation.

With the classical or quantized expressions of electrical fields in Eq. (A1-A4), we can obtain the Hamiltonian that describes the nonlinear interactions between pump, signal and idler fields. To express polarization directions of "TE" and "TM" plane waves with the same wave vector ($\hat{\mathbf{e}}_{s,0}$, $\hat{\mathbf{e}}_{s,1}$, $\hat{\mathbf{e}}_{i,0}$ and $\hat{\mathbf{e}}_{i,1}$), more simply and make the subsequent calculations easier, we introduce two angular variables to describe the wave vector direction of signal (idler) field. They are shown in Fig. 1 (b). One is $\varphi_{ks}$ ($\varphi_{ki}$), which is the angle between the transverse wave vector of signal (idler) field in x-y plane and x axis. Hence, $k_{sx} = k_{s\parallel} \cos\varphi_{ks}$ and $k_{sy} = k_{s\parallel} \sin\varphi_{ks}$ ($k_{ix} = k_{i\parallel} \cos\varphi_{ki}$ and $k_{iy} = k_{s\parallel} \sin\varphi_{ks}$). The other is $\theta_s(\omega_s, k_{s\parallel}, j_s)$ ($\theta_i(\omega_i, k_{i\parallel}, j_i)$) with $j_s = 1, 2$ ($j_i = 1, 2$), which is the angle between z axis and the wavevector. It can be calculated by $\sin\theta_{s,i}(\omega_{s,i}, k_{s,i\parallel}, j_{s,i}) = (-1)^{j_{s,i}} k_{s,i\parallel} / k_{s,i}(\omega_{s,i})$. With these variables, we can express the unit vectors $\hat{\mathbf{e}}_{s,0}$, $\hat{\mathbf{e}}_{s,1}$, $\hat{\mathbf{e}}_{i,0}$ and $\hat{\mathbf{e}}_{i,1}$ as

$$\hat{\mathbf{e}}_{s,0} = -\sin\varphi_{ks}\hat{\mathbf{e}}_x + \cos\varphi_{ks}\hat{\mathbf{e}}_y$$
$$\hat{\mathbf{e}}_{i,0} = -\sin\varphi_{ki}\hat{\mathbf{e}}_x + \cos\varphi_{ki}\hat{\mathbf{e}}_y$$
$$\hat{\mathbf{e}}_{s,1} = \sin\theta_s\hat{\mathbf{e}}_z - \cos\theta_s\cos\varphi_{ks}\hat{\mathbf{e}}_x - \cos\theta_s\sin\varphi_{ks}\hat{\mathbf{e}}_y$$
$$\hat{\mathbf{e}}_{i,1} = \sin\theta_i\hat{\mathbf{e}}_z - \cos\theta_i\cos\varphi_{ki}\hat{\mathbf{e}}_x - \cos\theta_i\sin\varphi_{ki}\hat{\mathbf{e}}_y. \tag{A5}$$

The different polarization components of pump, signal and idler fields are coupled by different components in the nonlinear susceptibility tensor in the SpFWM. Considering that the polarizations of the pump fields are along x directions and assume that the materials has spatial inversion symmetry, only the tensor components $\chi_{yyxx}^{(3)}$ and $\chi_{xxxx}^{(3)}$ are effective in the SpFWM process. We can introduce a ratio $r_\chi = \chi_{yyxx}^{(3)} / \chi_{xxxx}^{(3)}$ to describe the anisotropy of the nonlinear susceptibility in the ultrathin film. Then, the positive frequency part of the idler polarizations can be obtained as

$$\hat{\mathbf{P}}_i^{(+)}(x,y,z,t) = \varepsilon_0 \boldsymbol{\chi}^{(3)} \hat{\mathbf{x}}\hat{\mathbf{x}} [E_p^{(+)}(t,x,y,z)]^2 \sum_{j_s=1,2} \sum_{l_s=0,1} \hat{\mathbf{e}}_{s,l_s} \hat{E}_{s,j,l}^{(-)}(t,x,y,z) = -i(2\pi)^{-\frac{9}{2}} \int_{-\infty}^{\infty} d\omega_{p1} \int_{-\infty}^{\infty} d\omega_{p2} \int_{-\infty}^{\infty} d\omega_s \int_{-\infty}^{\infty} dk_{sx} \int_{-\infty}^{\infty} dk_{sy}$$

$$\times \int_{-\infty}^{\infty} dk_{p1x} \int_{-\infty}^{\infty} dk_{p1y} \int_{-\infty}^{\infty} dk_{p2x} \int_{-\infty}^{\infty} dk_{p2y} [\hat{\mathbf{x}}\hat{P}_{ix}^{(+)}(x,y,z,t) + \hat{\mathbf{y}}\hat{P}_{iy}^{(+)}(x,y,z,t)], \tag{A6}$$

where

$$\hat{P}_{ix}^{(+)}(x,y,z,t) = -\varepsilon_0 \chi_{xxxx} e^{i[(k_{p1x}+k_{p2x}-k_{sx})x+(k_{p1y}+k_{p2y}-k_{sy})y-(\omega_{p1}+\omega_{p2}-\omega_s)t]} E_p^{(+)}(\omega_{p1}, k_{p1x}, k_{p1y}, z) E_p^{(+)}(\omega_{p2}, k_{p2x}, k_{p2y}, z)$$

$$\times \sum_{j_s=0,1} [\sin\varphi_{ks} \hat{a}_{s,j,s}^\dagger(\omega_s, k_{sx}, k_{sy}, j_s, 0) + \cos\theta_{ks}\cos\varphi_{ks} \hat{a}_{s,j,p}^\dagger(\omega_s, k_{sx}, k_{sy}, j_s, 1)] e^{i[k_{p1z}+k_{p2z}-(-1)^{j_s} k_{sz}]z}, \tag{A7}$$

and

$$\hat{P}_{iy}^{(+)}(x,y,z,t) = \varepsilon_0 \chi_{yyxx} e^{i[(k_{sx}-k_{p1x}-k_{p2x})x+(k_{sy}-k_{p1y}-k_{p2y})y-(\omega_s-\omega_{p1}-\omega_{p2})t]} E_p^{(+)}(\omega_{p1}, k_{p1x}, k_{p1y}, z) E_p^{(+)}(\omega_{p2}, k_{p2x}, k_{p2y}, z)$$

$$\times \sum_{j_s=1,2} [\cos\varphi_{ks} \hat{a}_s^\dagger(\omega_s, k_{sx}, k_{sy}, j_s, 0) - \cos\theta_{ks}\sin\varphi_{ks} \hat{a}_{s,j,p}^\dagger(\omega_s, k_{sx}, k_{sy}, j_s, 0)] e^{i[k_{p1z}+k_{p2z}-(-1)^{j_s} k_{sz}]z}, \tag{A8}$$

Then, the interaction Hamiltonian for the FWM parametric process takes the form [2] of

$$\hat{H}_I = \int_{-L_x/2}^{L_x/2} dx \int_{-L_y/2}^{L_y/2} dy \int_{-d/2}^{d/2} dz [\hat{\mathbf{P}}_i^{(+)}(x,y,z,t) * \hat{\mathbf{E}}_i^{(-)}(t,x,y,z) + \text{H.c}] , \tag{A9}$$

where "*" presents the dot product. In Eq. (A8), $L_x$ ($L_y$) are the size of the film along $x$ ($y$) direction. Substituting Eqs. (A3), (A4) and (A5) into (A8) and implementing the integrals with variables $x$, $y$ and $z$, we can get the explicit expression of $\hat{H}_I$ as

$$\hat{H}_I = -\hbar \chi_{xxxx}^{(3)} \frac{(\tau_p r_p^2)^2}{2c} (2\pi)^{-6} E_0^2 \sum_{j_s=0,1} \sum_{j_i=0,1} \sum_{l_s=0,1} \sum_{l_i=0,1} \int d\omega_s \int d\omega_i \int d\omega_{p1} \int d\omega_{p2} \sqrt{\frac{\omega_s \omega_i}{n_s n_i}} e^{-i\Delta\omega t} e^{-\frac{\tau_p^2[(\omega_{p1}-\omega_{p0})^2+(\omega_{p2}-\omega_{p0})^2]}{2}}$$

$$\times \int dk_{p1x} \int dk_{p1y} \int dk_{p2x} \int dk_{p2y} \int dk_{sx} \int dk_{sy} \int dk_{ix} \int dk_{iy} L_x L_y d \operatorname{sinc}(\Delta k_x L_x/2) \operatorname{sinc}(\Delta k_y L_y/2)$$

$$\times \operatorname{sinc}\{[k_{p1}+k_{p2}-(-1)^{j_s} k_{sz}-(-1)^{j_i} k_{iz}]d/2\} e^{-\frac{r_p^2[(k_{p1x}^2+k_{p1y}^2)+(k_{p2x}^2+k_{p2y}^2)]}{2}} \cos^{l_s}[\theta_s(\omega_s,k_{s\|},j_s)] \cos^{l_i}[\theta_i(\omega_i,k_{i\|},j_i)]$$

$$\times \sum_{m=0,1} (r_\chi)^m \cos(\varphi_{ks} + \frac{l_s-m}{2}\pi) \cos(\varphi_{ki} + \frac{l_i-m}{2}\pi) \hat{a}_s^\dagger(\omega_s,k_{sx},k_{sy},j_s,l_s) \hat{a}_i^\dagger(\omega_i,k_{ix},k_{iy},j_i,l_i) + \text{H.c} , \tag{A10}$$

where $\Delta\omega = \omega_{p1}+\omega_{p2}-\omega_s-\omega_i$, $\Delta k_x = k_{p1x}+k_{p2x}-k_{sx}-k_{ix}$ and $\Delta k_y = k_{p1y}+k_{p2y}-k_{sy}-k_{iy}$. To simplify the calculation, the dependence of $\chi_{xxxx}^{(3)}$ on the angular frequencies of pump, signal and idler fields are neglected in the derivation of Eq. (A10). Similar with the treatment for SPDC [38], we consider that the size of the ultrathin film in the x-y plane is much larger than the transverse size of pump beam, and the terms $L_x \operatorname{sinc}(\Delta k_x L_x/2)$ and $L_y \operatorname{sinc}(\Delta k_y L_y/2)$ can be replaced by $2\pi\delta(\Delta k_x)$ and $2\pi\delta(\Delta k_y)$. Also, since the film is ultrathin along $z$ axis, the term $\operatorname{sinc}\{[k_{p1}+k_{p2}-(-1)^{j_s} k_{sz}-(-1)^{j_i} k_{iz}]d/2\}$ can be approximated by 1, i.e., the longitudinal phase mismatch can be neglected. Under the two approximations, the integrals about variables $k_{p1x}$, $k_{p1y}$, $k_{p2x}$ and $k_{p2y}$ in Eq. (A10) can be carried out and a simplified expression can be obtained as

$$\hat{H}_I = -[(2\pi)^{-3} n_p/6] \gamma P_p d\tau_p^2 r_p^2 \sum_{j_s=0,1} \sum_{j_i=0,1} \sum_{l_s=0,1} \sum_{l_i=0,1} \int d\omega_s \int d\omega_i \int d\omega_{p1} \int d\omega_{p2} \sqrt{\frac{\omega_s \omega_i}{n_s n_i}} e^{-\frac{\tau_p^2[(\omega_{p1}-\omega_{p0})^2+(\omega_{p2}-\omega_{p0})^2]}{2}}$$

$$\times \int dk_{sx} \int dk_{sy} \int dk_{ix} \int dk_{iy} e^{-\frac{r_p^2[(k_{sx}-k_{ix})^2+(k_{sy}-k_{iy})^2]}{4}} \cos^{l_s}[\theta_s(\omega_s,k_{s\|},j_s)] \cos^{l_i}[\theta_i(\omega_i,k_{i\|},j_i)]$$

$$\times \sum_{m=0,1} (r_\chi)^m \cos(\varphi_{ks} + \frac{l_s-m}{2}\pi) \cos(\varphi_{ki} + \frac{l_i-m}{2}\pi)$$

$$\times \hat{a}_s^\dagger(\omega_s,k_{sx},k_{sy},j_s,l_s) \hat{a}_i^\dagger(\omega_i,k_{ix},k_{iy},j_i,l_i) + \text{H.c} \tag{A11}$$

where $\gamma = n_2 k_{p0}/(\pi r_p^2)$ with $k_{p0} = \omega_{p0}/c$ is the nonlinear coefficient in the SpFWM, and $P_p = 2(\pi r_p^2) n_{p0} \varepsilon_0 c E_0^2$ is the peak power of the pump pulses. Here, $n_p$ is the linear refractive index of the film at the central frequency of the pulsed pump light. In the expression of $P_p$, $E_0$ is assumed to be real to simplify the calculations.

The initial states of the signal and idler fields are vacuum states and they evolves according to Schrödinger equation. For SpFWM, a perturbative solution of this equation is $|\psi\rangle_{s-i} \approx -(i/\hbar)\int_{-\infty}^{\infty} dt \hat{H}_I |0\rangle_s |0\rangle_i$ [9]. Substituting Eq. (A10) into this solution, we can get

$$|\psi\rangle_{s-i} \approx \beta \gamma P_p d(\pi r_p^2)(\sqrt{\pi} \tau_p) \sum_{j_s=0,1} \sum_{j_i=0,1} \sum_{l_s=0,1} \sum_{l_i=0,1} \int d\omega_s \int d\omega_i e^{-\frac{\tau_p^2(\omega_s+\omega_i-2\omega_{p0})^2}{4}}$$

$$\times \int dk_{sx} \int dk_{sy} \int dk_{ix} \int dk_{iy} e^{-\frac{r_p^2[(k_{sx}-k_{ix})^2+(k_{sy}-k_{iy})^2]}{4}} \cos^{l_s}[\theta_s(\omega_s,k_{s\|},j_s)] \cos^{l_i}[\theta_i(\omega_i,k_{i\|},j_i)]$$

$$\times \sum_{m=0,1} (r_\chi)^m \cos(\varphi_{ks} + \frac{l_s-m}{2}\pi) \cos(\varphi_{ki} + \frac{l_i-m}{2}\pi) \hat{a}_s^\dagger(\omega_s,k_{sx},k_{sy},j_s,l_s) \hat{a}_i^\dagger(\omega_i,k_{ix},k_{iy},j_i,l_i) |0\rangle_s |0\rangle_i . \tag{A12}$$

where the parameter $\beta = i(2\pi)^{-3} n_{p0}/3\sqrt{n_s n_i}$. In the derivation of Eq. (A11), $|\omega_{s,i}-\omega_{p0}| \ll \omega_{p0}$ is assumed. The Eq. (A11) can be written in a compact form as

$$|\psi\rangle = \sum_{j_s=0,1} \sum_{j_i=0,1} \sum_{l_s=0,1} \sum_{l_i=0,1} \int d^6 \mathbf{S} \psi(\mathbf{S},j_s,l_s,j_i,l_i) \hat{a}_s^\dagger(\mathbf{S}_s,j_s,l_s) \hat{a}_i^\dagger(\mathbf{S}_i,j_i,l_i) |0\rangle_s |0\rangle_i , \tag{A13}$$

which is the Eq. (1) in Sec. II. The meanings of $\mathbf{S}_{s,i}$ and $\mathbf{k}_{s\|,i\|} = [k_{sx,ix},k_{sy,iy}]$ can be found in the explanation of Eq. (1) in Sec. II. $\psi(\mathbf{S},j_s,l_s,j_i,l_i)$ is the biphoton probability amplitude function and its explicit expression is shown in Eq.

(2) in Sec. II.

## APPENDIX B: CALCULATION OF CORRELATION FUCNTION

In the case that only the photons propagating upwardly with respect to *x-y* plane are collected, two spatiotemporally varying annihilation operators can be constructed as

$$\hat{a}_s(t_s,\mathbf{r}_s) = (2\pi)^{-3/2} \sum_{j_s=0,1}\sum_{l_s=0,1} \int d\Omega_s \int d^2\mathbf{k}_{s\|} \hat{a}_s(\Omega_s,\mathbf{k}_{s\|},0,l_s) F_\Omega(\Omega_s) F_k(k_{s\|}) e^{i(\mathbf{k}_{s\|}\cdot\mathbf{r}_s - \Omega_s t_s)}, \quad (B1)$$

$$\hat{a}_i(t_i,\mathbf{r}_i) = (2\pi)^{-3/2} \sum_{j_i=0,1}\sum_{l_i=0,1} \int d\Omega_i \int d^2\mathbf{k}_{i\|} \hat{a}_i(\Omega_i,\mathbf{k}_{i\|},0,l_i) F_\Omega(\Omega_i) F_k(k_{i\|}) e^{i(\mathbf{k}_{i\|}\cdot\mathbf{r}_i - \Omega_i t_i)}, \quad (B2)$$

where $F_\Omega(\Omega_{s,i})$ and $F_k(k_{s\|,i\|})$ are the frequency and spatial frequency filtering functions in the collection and detection of signal and idler photons. The explicit expression of the filtering functions used in our calculations are shown in the Eqs. (6.1) and (6.2) in Sec. III.

The correlation function can be calculated by [31]

$$C(t_s,\mathbf{r}_s,t_i,\mathbf{r}_i) = \langle 0|\langle 0|\hat{a}_s(t_1,\mathbf{r}_1)\hat{a}_i(t_2,\mathbf{r}_2)|\psi\rangle_{s-i}. \quad (B3)$$

After substituting Eqs. (A12), (B1) and (B2) into (B3), and utilizing the commutation relationships

$$[\hat{a}_s(\Omega'_s,\mathbf{k}'_{s\|},j'_s,l'_s),\hat{a}_i^\dagger(\Omega_s,\mathbf{k}_{s\|},j_s,l_s)] = \delta(\Omega'_s-\Omega_s)\delta(\mathbf{k}'_{s\|}-\mathbf{k}_{s\|})\delta_{j'_s j_s}\delta_{l_s l'_s}, \quad (B4)$$

$$[\hat{a}_i(\Omega'_i,\mathbf{k}'_{i\|},j'_i,l'_i),\hat{a}_i^\dagger(\Omega_i,\mathbf{k}_{i\|},j_i,l_i)] = \delta(\Omega'_i-\Omega_i)\delta(\mathbf{k}'_{i\|}-\mathbf{k}_{i\|})\delta_{j'_i j_i}\delta_{l_i l'_i}, \quad (B5)$$

$$[\hat{a}_s(\Omega'_s,\mathbf{k}'_{s\|},j'_s,l'_s),\hat{a}_i^\dagger(\Omega_i,\mathbf{k}_{i\|},j_i,l_i)] = [\hat{a}_i(\Omega_i,\mathbf{k}_{i\|},j_i,l_i),\hat{a}_s^\dagger(\Omega'_s,\mathbf{k}'_{s\|},j'_s,l'_s)] = 0, \quad (B6)$$

we get the expression of $C(t_s,\mathbf{r}_s,t_i,\mathbf{r}_i)$ as

$$C(t_s,t_i,\mathbf{r}_s,\mathbf{r}_i) = \langle 0|\langle 0|\hat{a}_s(t_s,\mathbf{r}_s)\hat{a}_i(t_i,\mathbf{r}_i)|\psi\rangle$$

$$=(2\pi)^{-3}\sum_{l_s=0,1}\sum_{l_i=0,1}\int d^6\mathbf{S}'\psi(\mathbf{S},0,l_s,0,l_i)F_\Omega(\Omega_s)F_\Omega(\Omega_i)F_k(k_{s\|})F_k(k_{i\|})e^{-i(\Omega_s t_s+\Omega_i t_i)}e^{i(\mathbf{k}_{s\|}\mathbf{r}_s+\mathbf{k}_{i\|}\mathbf{r}_i)}. \quad (B7)$$

After substituting Eq. (2) into Eq. (B7), we can get

$$C(t_s,t_i,\mathbf{r}_s,\mathbf{r}_i) = C_c \sum_{l_s=0,1}\sum_{l_i=0,1}\int d\Omega_s \int d\Omega_i \int dk_{s\|}\int dk_{i\|} \int d\varphi_{ks}\int d\varphi_{ki} G_\Omega(\Omega_s,\Omega_i) G_k(\mathbf{k}_{s\|},\mathbf{k}_{i\|})\Phi(\varphi_{ks},\varphi_{ki},l_s,l_i)$$

$$\times\Theta(\Omega_s,k_{s\|},0,l_s)\Theta(\Omega_i,k_{i\|},0,l_i)U(\Omega_s,k_{s\|})U(\Omega_i,k_{i\|})F_\Omega(\Omega_s)F_\Omega(\Omega_i)F_k(k_{s\|})F_k(k_{i\|})e^{-i(\Omega_s t_s+\Omega_i t_i)}e^{i(\mathbf{k}_{s\|}\mathbf{r}_s+\mathbf{k}_{i\|}\mathbf{r}_i)}. \quad (B8)$$

where $C_c = (2\pi)^{-3}\beta\gamma P_p d_{eff}(\pi r_p^2)(\sqrt{\pi}\tau_p)$. In this expression, the integrals about vector $\mathbf{k}_{s\|}$ ($\mathbf{k}_{i\|}$) are implemented by using $d^2\mathbf{k}_{s\|}=k_{s\|}dk_{s\|}d\varphi_{ks}$.

The expression of $C(t_s,t_i,\mathbf{r}_s,\mathbf{r}_i)$ in Eq. (B8) can be simplified further by implementing the integrals with variables of $\varphi_{ks}$ and $\varphi_{ki}$. Considering the following equations

$$\mathbf{k}_{s\|}\cdot\mathbf{r}_s = k_{s\|}r_s\cos\varphi_s\cos\varphi_{ks} + k_{s\|}r_s\sin\varphi_s\sin\varphi_{ks} = k_{s\|}r_s\cos(\varphi_{ks}-\varphi_s), \quad (B9)$$

$$\mathbf{k}_{i\|}\cdot\mathbf{r}_i = k_{i\|}r_i\cos\varphi_i\cos\varphi_{ki} + k_{i\|}r_i\sin\varphi_i\sin\varphi_{ki} = k_{i\|}r_i\cos(\varphi_{ki}-\varphi_i), \quad (B10)$$

$$e^{ik_{s,i\|}r_{s,i}\cos(\varphi_{s,i}-\varphi_{ks,i})} = \sum_{n=-\infty}^{\infty} i^n J_n(k_{s,i\|}r_{s,i})e^{in(\varphi_{s,i}-\varphi_{ks,i})}, \quad (B11)$$

$$e^{\frac{r_p^2 k_{s\|}k_{i\|}}{2}\cos(\varphi_{ks}-\varphi_{ki})} = \sum_{n=-\infty}^{\infty}(-1)^n I_n(\frac{r_p^2 k_{s\|}k_{i\|}}{2})e^{in(\varphi_{ks}-\varphi_{ki})}, \quad (B12)$$

we can finish the integrals about $\varphi_{ks}$ and $\varphi_{ki}$ in Eq. (B8) and get

$$C(\Delta t,\Delta r) = C_c \sum_{l_s=0,1}\sum_{l_i=0,1}\int d\omega_s \int d\omega_i G_\Omega(\Omega_s,\Omega_i) F_\Omega(\Omega_s) F_\Omega(\Omega_i) e^{-i(\Omega_i t_i + \Omega_s t_s)}$$

$$\times \int k_{s\|}dk_{s\|}\int k_{i\|}dk_{i\|} g_k(k_{s\|},k_{i\|}) J(k_{i\|},\Delta r,\varphi_i)$$

$$\times\Theta(\Omega_s,k_{s\|},0,l_s)\Theta(\Omega_i,k_{s\|},0,l_i)U(\Omega_s,k_{s\|})U(\Omega_i,k_{i\|}), \quad (B13)$$

where $g_k(k_{s\|},k_{i\|}) = e^{-r_p^2(k_{s\|}^2+k_{i\|}^2)/4}$. The expression of $J(k_{i\|},r_i,\varphi_i)$ is

$$J(k_{i\|},r_i,\varphi_i) = \frac{1}{4}\sum_{n=-\infty}^{\infty} I_n(\frac{r_p^2 k_{s\|}k_{i\|}}{2})\{J_{n+1}(k_{s\|}r_s)e^{i(n+1)\varphi_s}[(1-r_\chi)J_{n-1}(k_{i\|}r_i)e^{-i(n-1)\varphi_i}e^{i(\frac{l_s+l_i}{2}\pi)}$$

$$-(1+r_\chi)J_{n+1}(k_{i\|}r_i)e^{-i(n+1)\varphi_i}e^{i(\frac{l_s-l_i}{2}\pi)}]+J_{n-1}(k_{s\|}r_s)e^{i(n-1)\varphi_s}[(1-r_\chi)J_{n+1}(k_{i\|}r_i)e^{-i(n+1)\varphi_i}e^{-i(\frac{l_s+l_i}{2}\pi)}$$

$$-(1+r_\chi)J_{n-1}(k_{i\|}r_i)e^{-i(n-1)\varphi_i}e^{-i(\frac{l_s-l_i}{2}\pi)}]\}, \tag{B14}$$

where $I_n(*)$, $J_{n+1}(*)$, and $J_{n-1}(*)$ are the $n^{th}$-order modified Bessel function, $(n+1)^{th}$-order Bessel function, and $(n-1)^{th}$-order Bessel function, respectively.

When $\mathbf{r}_s$ is set at origin, i.e., $r_s=0$, Eq. (14) is simplified to

$$J(k_{i\|},r_i,\varphi_i)=\frac{1}{2}\{(1+r_\chi)J_0(k_{i\|}r_i)\cos[\frac{\pi}{2}(l_s-l_i)]-(1-r_\chi)J_2(k_{i\|}r_i)\cos[\frac{\pi}{2}(l_s+l_i)+2\varphi_i]\}. \tag{B15}$$

Equations (B13) and (B15) are the Eqs. (7) and (8) in Sec. III, respectively.

## APPENDIX C: CALCULATION OF PURITY OF REDUCED DENSITY MATRIX

With the collection and detection configuration in the derivation of Eq. (5), we can get an effective biphoton probability amplitude function as [21]

$$\psi_e(\Omega_s,\Omega_i,\mathbf{k}_{s\|},\mathbf{k}_{i\|},l_s,l_i)=\psi(\Omega_s,\Omega_i,\mathbf{k}_{s\|},\mathbf{k}_{i\|},0,l_s,0,l_i)F_\Omega(\Omega_s)F_\Omega(\Omega_i)F_k(\mathbf{k}_{s\|})F_k(\mathbf{k}_{i\|}), \tag{C1}$$

where the filtering functions $F_\Omega(\Omega_s)$, $F_\Omega(\Omega_i)$, $F_k(\mathbf{k}_{s\|})$ and $F_k(\mathbf{k}_{i\|})$ are the same to those in Eqs. (6.1) and (6.2). Then an effective biphoton state can be written as

$$|\psi\rangle_e=\sum_{l_s=0,1}\sum_{l_i=0,1}\int d\Omega_s\int d\Omega_i\int d\mathbf{k}_{s\|}\int d\mathbf{k}_{i\|}\psi_e(\Omega_s,\Omega_i,\mathbf{k}_{s\|},\mathbf{k}_{i\|},l_s,l_i)a_s^\dagger(\Omega_s,\mathbf{k}_{s\|},l_s)a_i^\dagger(\Omega_i,\mathbf{k}_{i\|},l_i)|0\rangle_s|0\rangle_i. \tag{C2}$$

To get its density matrix, this state should be normalized [11]. The normalized constant $N_a$ would be obtained by

$$N_a=[\sum_{l_s=0,1}\sum_{l_i=0,1}\int d\Omega_s\int d\Omega_i\int d\mathbf{k}_{s\|}\int d\mathbf{k}_{i\|}|\psi_e(\Omega_s,\Omega_i,\mathbf{k}_{s\|},\mathbf{k}_{i\|},l_s,l_i)|^2]^{-1/2}|. \tag{C3}$$

The part within the bracket in Eq. (C3) can be expanded into

$$\sum_{l_s=0,1}\sum_{l_i=0,1}\int d\Omega_s\int d\Omega_i\int d\mathbf{k}_{s\|}\int d\mathbf{k}_{i\|}|\psi_e(\Omega_s,\Omega_i,\mathbf{k}_{s\|},\mathbf{k}_{i\|},l_s,l_i)|^2$$

$$=C_0^2\sum_{l_s=0,1}\sum_{l_i=0,1}\int d\Omega_s\int d\Omega_i\int dk_{s\|}\int dk_{i\|}\int d\varphi_{ks}\int d\varphi_{ki}G_\Omega^2(\Omega_s,\Omega_i)e^{-\frac{r_p^2[k_{s\|}^2+k_{i\|}^2+2k_{s\|}k_{i\|}\cos(\varphi_{ks}-\varphi_{ki})]}{2}}$$

$$\times[r_\chi\sin(\varphi_{ks}+\frac{l_s\pi}{2})\sin(\varphi_{ki}+\frac{l_i\pi}{2})+\cos(\varphi_{ks}+\frac{l_s\pi}{2})\cos(\varphi_{ki}+\frac{l_i\pi}{2})]^2$$

$$\times\Theta^2(\Omega_s,k_{s\|},0,l_s)\Theta^2(\Omega_i,k_{i\|},0,l_i)U(\Omega_s,k_{s\|})U(\Omega_i,k_{i\|}). \tag{C4}$$

where $C_0=|\beta|(\gamma P_p d_{eff})(\pi r_p^2)(\sqrt{\pi}\tau_p)$. With the equation

$$e^{-r_p^2 k_{i\|}k_{s\|}\cos(\varphi_{ks}-\varphi_{ki})}=\sum_{n=-\infty}^{\infty}(-1)^n I_n(r_p^2 k_{i\|}k_{s\|})e^{in(\varphi_{ks}-\varphi_{ki})}, \tag{C5}$$

Eq. (C4) can be simplified. Specifically, after Eq. (C5) is substituted into Eq. (C4) and the integrals about $\varphi_{ks}$ and $\varphi_{ki}$ are implemented, we get

$$\sum_{l_s=0,1}\sum_{l_i=0,1}\int d\Omega_s\int d\Omega_i\int d\mathbf{k}_{s\|}\int d\mathbf{k}_{i\|}|\psi_e(\Omega_s,\Omega_i,\mathbf{k}_{s\|},\mathbf{k}_{i\|},l_s,l_i)|^2$$

$$=\frac{1}{4}\sum_{l_s=0,1}\sum_{l_i=0,1}\int_{-\infty}^{\infty}d\Omega_s\int_{-\infty}^{\infty}d\Omega_i\int k_{s\|}dk_{s\|}\int k_{i\|}dk_{i\|}$$

$$\times G_\Omega^2(\Omega_s,\Omega_i)g_k^2(k_{s\|},k_{i\|})F_\Omega^2(\Omega_s)F_\Omega^2(\Omega_i)F_k^2(k_{s\|})F_k^2(k_{i\|})$$

$$\times I(k_{s\|},k_{i\|},l_s,l_i)\Theta^2(\Omega_s,k_{s\|},0,l_s)\Theta^2(\Omega_i,k_{i\|},0,l_i)U(\Omega_s,k_{s\|})U(\Omega_i,k_{i\|}), \tag{C6}$$

where $I(k_{s\|},k_{i\|},l_s,l_i)=(1+r_\chi^2)I_0(r_p^2 k_{s\|}k_{i\|})+(1+r_\chi)^2\cos[(l_s-l_i)\pi]I_2(r_p^2 k_{s\|}k_{i\|})$. With Eqs. (C3-C6), $N_a$ can be numerically calculated.

The normalized effective biphoton state is

$$|\psi\rangle_{Ne}=N_a|\psi\rangle_e. \tag{C7}$$

The density matrix of this state is

$$\rho = |N_a|^2 \sum_{l_{s1}=0,1} \sum_{l_{s2}=0,1} \sum_{l_{i1}=0,1} \sum_{l_{i2}=0,1} \int_{-\infty}^{\infty} d\Omega_{s1} \int_{-\infty}^{\infty} d\Omega_{i1} \int_{-\infty}^{\infty} d\Omega_{s2} \int_{-\infty}^{\infty} d\Omega_{i2} \int d^2\mathbf{k}_{s1\|} \int d^2\mathbf{k}_{i1\|} \int d^2\mathbf{k}_{s2\|} \int d^2\mathbf{k}_{i2\|}$$
$$\times \psi_e(\Omega_{s1},\Omega_{i1},\mathbf{k}_{s1\|},\mathbf{k}_{i1\|},l_{s1},l_{i1}) \psi_e^*(\Omega_{s2},\Omega_{i2},\mathbf{k}_{s2\|},\mathbf{k}_{i2\|},l_{s2},l_{i2})$$
$$\times a_s^\dagger(\Omega_{s2},\mathbf{k}_{s2\|},l_{s2}) a_i^\dagger(\Omega_{i2},\mathbf{k}_{i2\|},l_{i2}) a_s(\Omega_{s1},\mathbf{k}_{s1\|},l_{s1}) a_i(\Omega_{i1},\mathbf{k}_{i1\|},l_{i1}) \,. \tag{C8}$$

Tracing out the parameters $\mathbf{k}_{s1,2\|}$, $\mathbf{k}_{i1,2\|}$, $l_{s1,2}$ and $l_{i1,2}$ can reduce the density matrix $\rho$ to $\rho_{\omega_s,\omega_i}$ which describes the temporal biphoton state[21]. This process determines

$$\rho_{\omega_s,\omega_i} = \sum_{l_s=0,1} \sum_{l_i=0,1} \int d\mathbf{k}_{s\|} \int d\mathbf{k}_{i\|} a_s(\mathbf{k}_{s\|},l_s) a_i(\mathbf{k}_{i\|},l_i) \rho a_s^\dagger(\mathbf{k}_{s\|},l_s) a_i^\dagger(\mathbf{k}_{i\|},l_i) \,. \tag{C9}$$

After Eq. (C8) is substituted into Eq. (C9) and the commutations

$$a_s(\Omega_{s,i},\mathbf{k}_{s\|,i\|},l_{s,i}) a_{s,i}^\dagger(\mathbf{k}'_{s,i\|},l'_{s,i}) = a_{s,i}(\Omega_{s,i}) \delta(\mathbf{k}_{s,i\|}-\mathbf{k}'_{s,i\|}) \delta_{l_{s,i},l'_{s,i}} + a_{s,i}^\dagger(\mathbf{k}'_{s,i\|},l'_{s,i}) a_{s,i}(\Omega_{s,i},\mathbf{k}_{s,i\|},l_{s,i}) \,,$$
$$a_{s,i}(\mathbf{k}'_{s,i\|},l'_{s,i}) a_{s,i}^\dagger(\Omega_{s,i},\mathbf{k}_{s\|,i\|},l_{s,i}) = a_{s,i}^\dagger(\Omega_{s,i}) \delta(\mathbf{k}_{s,i\|}-\mathbf{k}'_{s,i\|}) \delta_{l_{s,i},l'_{s,i}} + a_{s,i}^\dagger(\mathbf{k}'_{s,i\|},l'_{s,i}) a_{s,i}(\Omega_{s,i},\mathbf{k}_{s,i\|},l_{s,i}) \tag{C10}$$

are used, the explicit expression of $\rho_{\omega_s,\omega_i}$ can be calculated as

$$\rho_{\omega_s,\omega_i} = |N_a|^2 \sum_{l_s=0,1} \sum_{l_i=0,1} \int_{-\infty}^{\infty} d\Omega_{s1} \int_{-\infty}^{\infty} d\Omega_{i1} \int_{-\infty}^{\infty} d\Omega_{s2} \int_{-\infty}^{\infty} d\Omega_{i2} \int d\mathbf{k}_{s\|} \int d\mathbf{k}_{i\|}$$
$$\times \psi_e(\Omega_{s1},\Omega_{i1},\mathbf{k}_{s\|},\mathbf{k}_{i\|},l_s,l_i) \psi_e^*(\Omega_{s2},\Omega_{i2},\mathbf{k}_{s\|},\mathbf{k}_{i\|},l_s,l_i) a_s^\dagger(\Omega_{s1}) a_i^\dagger(\Omega_{i1}) a_s(\Omega_{s2}) a_i(\Omega_{i2}) \,. \tag{C11}$$

Next, the purity $\mathbf{Tr}(\rho_{\omega_s,\omega_i}^2)$ can be obtained as

$$\mathbf{Tr}(\rho_{\omega_s,\omega_i}^2) = \int_{-\infty}^{\infty} d\Omega_s \int_{-\infty}^{\infty} d\Omega_i a_s(\Omega_s) a_i(\Omega_i) \rho_{\omega_s,\omega_i}^2 a_s^\dagger(\Omega_s) a_i^\dagger(\Omega_i)$$
$$= |N_a|^2 \sum_{l_{s1}=0,1} \sum_{l_{s2}=0,1} \sum_{l_{i1}=0,1} \sum_{l_{i2}=0,1} \int_{-\infty}^{\infty} d\Omega_{s1} \int_{-\infty}^{\infty} d\Omega_{i1} \int_{-\infty}^{\infty} d\Omega_{s2} \int_{-\infty}^{\infty} d\Omega_{i2} \int d^2\mathbf{k}_{s1\|} \int d^2\mathbf{k}_{i1\|} \int d^2\mathbf{k}_{s2\|} \int d^2\mathbf{k}_{i2\|}$$
$$\times \psi_e(\Omega_{s1},\Omega_{i1},\mathbf{k}_{s1\|},\mathbf{k}_{i1\|},l_{s1},l_{i1}) \psi_e^*(\Omega_{s1},\Omega_{i1},\mathbf{k}_{s2\|},\mathbf{k}_{i2\|},l_{s2},l_{i2})$$
$$\times \psi_e(\Omega_{s2},\Omega_{i2},\mathbf{k}_{s2\|},\mathbf{k}_{i2\|},l_{s2},l_{i2}) \psi_e^*(\Omega_{s2},\Omega_{i2},\mathbf{k}_{s1\|},\mathbf{k}_{i1\|},l_{s1},l_{i1}) \,. \tag{C12}$$

Then, after substituting Eqs. (2) and (C1) into Eq. (C12), we obtain

$$\mathbf{Tr}(\rho_{\omega_s,\omega_i}^2) = \frac{B}{|N|^2} \,, \tag{C13}$$

where $|N| = 1/|N_a|$ and

$$B = \frac{1}{16} \sum_{l_{s1}=0,1} \sum_{l_{s2}=0,1} \sum_{l_{i1}=0,1} \sum_{l_{i2}=0,1} \int_{-\infty}^{\infty} d\Omega_{s1} \int_{-\infty}^{\infty} d\Omega_{i1} \int_{-\infty}^{\infty} d\Omega_{s2} \int_{-\infty}^{\infty} d\Omega_{i2} \int k_{s1\|} dk_{s1\|} \int k_{s2\|} dk_{s2\|} \int k_{i1\|} dk_{i1\|} \int k_{i2\|} dk_{i2\|}$$
$$\times G_\Omega^2(\Omega_{s1},\Omega_{i1}) g_k^2(k_{s1\|},k_{i1\|}) F_\Omega^2(\Omega_{s1}) F_\Omega^2(\Omega_{i1}) F_k^2(k_{s1\|}) F_k^2(k_{i1\|}) I(k_{s1\|},k_{i1\|},l_{s1},l_{i1})$$
$$\times G_\Omega^2(\Omega_{s2},\Omega_{i2}) g_k^2(k_{s2\|},k_{i2\|}) F_\Omega^2(\Omega_{s2}) F_\Omega^2(\Omega_{i2}) F_k^2(k_{s2\|}) F_k^2(k_{i2\|}) I(k_{s2\|},k_{i2\|},l_{s2},l_{i2})$$
$$\times \Theta^2(\Omega_{s1},k_{s1\|},0,l_{s1}) \Theta^2(\Omega_{i1},k_{i1\|},0,l_{i1}) U(\Omega_{s1},k_{s1\|}) U(\Omega_{i1},k_{i1\|})$$
$$\times \Theta^2(\Omega_{s1},k_{s2\|},0,l_{s2}) \Theta^2(\Omega_{i1},k_{i2\|},0,l_{i2}) U(\Omega_{s1},k_{s2\|}) U(\Omega_{i1},k_{i2\|})$$
$$\times \Theta^2(\Omega_{s2},k_{s1\|},0,l_{s1}) \Theta^2(\Omega_{i2},k_{i1\|},0,l_{i1}) U(\Omega_{s2},k_{s1\|}) U(\Omega_{i2},k_{i1\|})$$
$$\times \Theta^2(\Omega_{s2},k_{s2\|},0,l_{s2}) \Theta^2(\Omega_{i2},k_{i2\|},0,l_{i2}) U(\Omega_{s2},k_{s2\|}) U(\Omega_{i2},k_{i2\|}) \,. \tag{C14}$$

The Eq. (C13) is the Eq. (11) in Sec. IV, while the Eq. (13) is a compact form of Eq. (C14).

## APPENDIX D: CALCULATION OF SPATIAL SCHMIDT NUMBER

To calculate the spatial Schmidt number, the frequency detunings of the signal and idler photons are fixed at $\Omega_s$ and $\Omega_i$, respectively. If only the upwardly propagating photons are collected, a spatial biphoton state can be obtained as

$$|\psi\rangle_k = \sum_{l_s=0,1} \sum_{l_i=0,1} \int d\mathbf{k}_{s\|} \int d\mathbf{k}_{i\|} \psi(\mathbf{k}_{s\|},\mathbf{k}_{i\|},l_s,l_i) a_s^\dagger(\mathbf{k}_{s\|},l_s) a_i^\dagger(\mathbf{k}_{i\|},l_i) |0\rangle_s |0\rangle_s \,, \tag{D1}$$

where $\psi(\mathbf{k}_{s\|},\mathbf{k}_{i\|},l_s,l_i)$ is the $\psi(\Omega_s,\Omega_i,\mathbf{k}_{s\|},\mathbf{k}_{i\|},j_s,l_s,j_i,l_i)$ in Eq. (3) with fixed $\Omega_s$, $\Omega_i$ and $j_{s,i}=1$. Before the next calculation, the state should be normalized, and the normalized factor $N_k$ can be obtained by

$$N_k = |[\sum_{l_s=0,1} \sum_{l_i=0,1} \int d\mathbf{k}_{s\|} \int d\mathbf{k}_{i\|} |\psi(\mathbf{k}_{s\|},\mathbf{k}_{i\|},l_s,l_i)|^2]^{-1/2}| \,, \tag{D2}$$

With the explicit expression of $\psi(\mathbf{k}_{s\|}, \mathbf{k}_{i\|}, l_s, l_i)$ obtained from Eq. (3), the denominator at the right hand of Eq. (D2) can be expressed as

$$\sum_{l_s=0,1}\sum_{l_i=0,1}\int d\mathbf{k}_{s\|}\int d\mathbf{k}_{i\|}|\psi(\mathbf{k}_{s\|},\mathbf{k}_{i\|},l_s,l_i)|^2 = \frac{1}{4}C_0^2 G_\Omega^2(\Omega_s,\Omega_i)\sum_{l_s=0,1}\sum_{l_i=0,1}\int k_{s\|}dk_{s\|}\int k_{i\|}dk_{i\|}g_k^2(k_{s\|},k_{i\|})$$
$$\times I(k_{s\|},k_{i\|},l_s,l_i)\Theta^2(\Omega_s,k_{s\|},0,l_s)\Theta^2(\Omega_i,k_{i\|},0,l_i)U(\Omega_s,k_{s\|})U(\Omega_i,k_{i\|}) . \quad (D3)$$

where $C_0$ is the same with that in Eq. (C4) in Appendix C. This expression can give the expression of density matrix of the normalized $|\psi\rangle_k$ as

$$\rho_k = |N_k|^2 \sum_{l_{s1}=0,1}\sum_{l_{i1}=0,1}\sum_{l_{s2}=0,1}\sum_{l_{i2}=0,1}\int d\mathbf{k}_{s1\|}\int d\mathbf{k}_{i1\|}\int d\mathbf{k}_{s2\|}\int d\mathbf{k}_{i2\|}$$
$$\times \psi(\mathbf{k}_{s1\|},\mathbf{k}_{i1\|},l_{s1},l_{i1})\psi^*(\mathbf{k}_{s1\|},\mathbf{k}_{i2\|},l_{s2},l_{i2})\psi(\mathbf{k}_{s2\|},\mathbf{k}_{i2\|},l_{s2},l_{i2})\psi^*(\mathbf{k}_{s1\|},\mathbf{k}_{i1\|},l_{s1},l_{i1})$$
$$\times a_s^\dagger(\mathbf{k}_{s1\|},l_s)a_i^\dagger(\mathbf{k}_{i1\|},l_i)a_s(\mathbf{k}_{s1\|},l_s)a_i(\mathbf{k}_{i1\|},l_i) . \quad (D4)$$

To calculate the spatial Schmidt number, the polarization degree of freedom should be traced out in the density matrix. Also, the transverse wave vector of either signal or idler field has to be traced out [11]. After the dimensions of $\mathbf{k}_{i\|}$, $l_s$ and $l_i$ in $\rho_k$ are traced out, a reduced density matrix $\rho_{\mathbf{k}_s}$ is obtained as

$$\rho_{\mathbf{k}_s} = |N_k|^2 \sum_{l_s=0,1}\sum_{l_i=0,1}\int d\mathbf{k}_{s1\|}\int d\mathbf{k}_{s2\|}\int d\mathbf{k}_{i\|}\psi(\mathbf{k}_{s1\|},\mathbf{k}_{i\|},l_s,l_i)\psi^*(\mathbf{k}_{s2\|},\mathbf{k}_{i\|},l_s,l_i)a_s^\dagger(\mathbf{k}_{s1\|})a_s(\mathbf{k}_{s2\|}) . \quad (D5)$$

The purity of $\rho_{\mathbf{k}_s}$ is $\mathbf{Tr}(\rho_{\mathbf{k}_s}^2)$ and can be calculated as

$$\mathbf{Tr}(\rho_{\mathbf{k}_s}^2) = \int d\mathbf{k}_{s\|}a_s(\mathbf{k}_{s\|})\rho_{\mathbf{k}_s}^2 a_s^\dagger(\mathbf{k}_{s\|}) . \quad (D6)$$

After Eq. (D5) is substituted into Eq. (D6), $\mathbf{Tr}(\rho_{\mathbf{k}_s}^2)$ can be expressed as

$$\mathbf{Tr}(\rho_{\mathbf{k}_s}^2) = \frac{T}{N_K^2}, \quad (D7)$$

where $N_K = 1/|N_k|^2$ and

$$T = \sum_{l_{s1}=0,1}\sum_{l_{i1}=0,1}\sum_{l_{s2}=0,1}\sum_{l_{i2}=0,1}\int d\mathbf{k}_{s1\|}\int d\mathbf{k}_{s2\|}\int d\mathbf{k}_{i1\|}\int d\mathbf{k}_{i2\|}$$
$$\times \psi(\mathbf{k}_{s1\|},\mathbf{k}_{i1\|},l_{s1},l_{i1})\psi^*(\mathbf{k}_{s2\|},\mathbf{k}_{i1\|},l_{s1},l_{i1})\psi(\mathbf{k}_{s2\|},\mathbf{k}_{i2\|},l_{s2},l_{i2})\psi^*(\mathbf{k}_{s1\|},\mathbf{k}_{i2\|},l_{s2},l_{i2}) . \quad (D8)$$

Then, it is easy to get the spatial Schmidt number [11] as

$$K_F = 1/\mathbf{Tr}(\rho_{\mathbf{k}_s}^2) = N_K^2/T . \quad (D9)$$

The explicit from of $\psi(\mathbf{k}_{s\|},\mathbf{k}_{i\|},l_s,l_i)$ can be obtained according to the expression of the biphoton probability amplitude function in Eq. (2). After $\psi(\mathbf{k}_{s\|},\mathbf{k}_{i\|},l_s,l_i)$ is substituted into (D8), and the integrals with variables of $\varphi_{ks}$ and $\varphi_{ki}$ are implemented, Eq. (D8) can be expressed as

$$T = \sum_{l_{s1}=0,1}\sum_{l_{i1}=0,1}\sum_{l_{s2}=0,1}\sum_{l_{i2}=0,1}\sum_{m_1=0,1}\sum_{m_2=0,1}\sum_{m_3=0,1}\sum_{m_4=0,1}\int k_{s1\|}dk_{s1\|}\int k_{s1\|}dk_{s2\|}\int k_{s1\|}dk_{i1\|}\int k_{s1\|}dk_{i2\|}$$
$$\times e^{-\frac{r_p^2(k_{s1\|}^2+k_{s2\|}^2+k_{i1\|}^2+k_{i2\|}^2)}{2}}(r_\chi)^{m_1+m_2+m_3+m_4}(F_1+F_2+F_3+F_4+F_5+F_6+F_7+F_8)$$
$$\times \Theta(\Omega_s,k_{s1\|},0,l_{s1})\Theta(\Omega_s,k_{s1\|},0,l_{s2})\Theta(\Omega_s,k_{s2\|},0,l_{s1})\Theta(\Omega_s,k_{s2\|},0,l_{s2})$$
$$\times \Theta(\Omega_i,k_{i1\|},0,2l_{i1})\Theta(\Omega_i,k_{i2\|},0,2l_{i2})U(\Omega_s,k_{s1\|})U(\Omega_s,k_{s2\|})U(\Omega_i,k_{i1\|})U(\Omega_i,k_{i2\|}) , \quad (D10)$$

where

$$F_1 = \frac{1}{16}\cos[\frac{(l_{s1}-l_{s2}+m_1-m_4)\pi}{2}]\cos[\frac{(l_{s1}-l_{s2}+m_2-m_3)\pi}{2}]\cos[\frac{(m_3-m_4)\pi}{2}]$$
$$\times \cos[\frac{(m_1-m_2)\pi}{2}]\sum_{n=-\infty}^{\infty}I_n(\frac{r_p^2 k_{s1\|}k_{i1\|}}{2})I_n(\frac{r_p^2 k_{s1\|}k_{i2\|}}{2})I_n(\frac{r_p^2 k_{s2\|}k_{i2\|}}{2})I_n(\frac{r_p^2 k_{s2\|}k_{i1\|}}{2}) , \quad (D11)$$

$$F_2 = \frac{1}{128}\cos[(l_{i1}-l_{i2}+m_3-m_1)\pi]$$

$$\times \sum_{n=-\infty}^{\infty} I_n(\frac{r_p^2 k_{s1\|}k_{i1\|}}{2})I_{n+2}(\frac{r_p^2 k_{s1\|}k_{i2\|}}{2})I_{n+4}(\frac{r_p^2 k_{s2\|}k_{i2\|}}{2})I_{n+2}(\frac{r_p^2 k_{s2\|}k_{i1\|}}{2}), \tag{D12}$$

$$F_3 = \frac{1}{128}\cos[(l_{s1}-l_{s2}-l_{i1}-l_{i2})\pi]$$

$$\times \sum_{n=-\infty}^{\infty} I_n(\frac{r_p^2 k_{s1\|}k_{i1\|}}{2})I_{n+2}(\frac{r_p^2 k_{s1\|}k_{i2\|}}{2})I_n(\frac{r_p^2 k_{s2\|}k_{i2\|}}{2})I_{n+2}(\frac{r_p^2 k_{s2\|}k_{i1\|}}{2}), \tag{D13}$$

$$F_4 = \frac{1}{128}\cos\{[(l_{s1}+l_{s2}+l_{i1}+l_{i2})-m_1-m_2-m_3-m_4]\pi\}$$

$$\times \sum_{n=-\infty}^{\infty} I_n(\frac{r_p^2 k_{s1\|}k_{i1\|}}{2})I_{n+2}(\frac{r_p^2 k_{s1\|}k_{i2\|}}{2})I_n(\frac{r_p^2 k_{s2\|}k_{i2\|}}{2})I_{n+2}(\frac{r_p^2 k_{s2\|}k_{i1\|}}{2}), \tag{D14}$$

$$F_5 = \frac{1}{32}\cos[\frac{(l_{s1}+l_{s2}+2l_{i2}-m_2-2m_3-m_4)\pi}{2}]$$

$$\times \sum_{n=-\infty}^{\infty} I_n(\frac{r_p^2 k_{s1\|}k_{i1\|}}{2})I_n(\frac{r_p^2 k_{s1\|}k_{i2\|}}{2})I_{n+2}(\frac{r_p^2 k_{s2\|}k_{i2\|}}{2})I_n(\frac{r_p^2 k_{s2\|}k_{i1\|}}{2}), \tag{D15}$$

$$F_6 = \frac{1}{32}\cos[\frac{(2l_{i1}-2l_{i2}-m_1-m_2+m_3+m_4)\pi}{2}]$$

$$\times \sum_{n=-\infty}^{\infty} I_n(\frac{r_p^2 k_{s1\|}k_{i1\|}}{2})I_{n+2}(\frac{r_p^2 k_{s1\|}k_{i2\|}}{2})I_{n+2}(\frac{r_p^2 k_{s2\|}k_{i2\|}}{2})I_n(\frac{r_p^2 k_{s2\|}k_{i1\|}}{2}), \tag{D16}$$

$$F_7 = \frac{1}{32}\cos[(\frac{2l_{s1}+2l_{s2}-m_1-m_4+m_2+m_3}{2})\pi]$$

$$\times \sum_{n=-\infty}^{\infty} I_n(\frac{r_p^2 k_{s1\|}k_{i1\|}}{2})I_n(\frac{r_p^2 k_{s1\|}k_{i2\|}}{2})I_{n+2}(\frac{r_p^2 k_{s2\|}k_{i2\|}}{2})I_{n+2}(\frac{r_p^2 k_{s2\|}k_{i1\|}}{2}), \tag{D17}$$

$$F_8 = \frac{1}{32}\cos[(\frac{l_{s1}+l_{s2}+2l_{i1}-2m_1-m_2-m_4}{2})\pi]$$

$$\times \sum_{n=-\infty}^{\infty} I_n(\frac{r_p^2 k_{s1\|}k_{i1\|}}{2})I_{n+2}(\frac{r_p^2 k_{s1\|}k_{i2\|}}{2})I_{n+2}(\frac{r_p^2 k_{s2\|}k_{i2\|}}{2})I_{n+2}(\frac{r_p^2 k_{s2\|}k_{i1\|}}{2}). \tag{D18}$$

Utilizing Eqs. (D2) and (D9-D18), the spatial Schmidt number of the biphoton state described by Eq. (2) can be calculated numerically. However, a simpler form of (D9) can be derived under two assumptions. Firstly, we only collect the photon pairs with "TE/TE" polarizations. Secondly, pump beam has relatively large transverse radius, so that the signal and idler photon pairs generated in the SpFWM have $\mathbf{k}_{s\|} \approx \mathbf{k}_{i\|}$. (With this assumption, the term of $G_k(\mathbf{k}_{s\|},\mathbf{k}_{i\|})=e^{-r_p^2(\mathbf{k}_{s\|}+\mathbf{k}_{i\|})^2/4}$ in Eq. (3) can be simplified as $4\pi\delta(\mathbf{k}_{s\|}+\mathbf{k}_{i\|})/r_p^2$). After these two assumptions, Eq. (D8) can be simplified to

$$T = (\frac{4\pi}{r_p^2})^3 \int d\mathbf{k}_{s\|} U(\Omega_s, k_{s\|})U(\Omega_i, k_{s\|}). \tag{D19}$$

According to $N_K = 1/|N_k|^2$ and Eq. (D2), a simplified form of $N_K$ can be obtained as

$$N_K = (\frac{2\pi}{r_p^2})^2[\int d\mathbf{k}_{s\|} U(\Omega_s, k_{s\|})U(\Omega_i, k_{s\|})]^2. \tag{D20}$$

With Eqs. (D19) and (D20), an approximated expression of $K_F$ can be obtained as

$$K_F = \frac{1}{8}\frac{\int d\mathbf{k}_{s\|} u(\Omega_s, \mathbf{k}_{s\|})u(\Omega_i, -\mathbf{k}_{s\|})}{\frac{2\pi}{r_p^2}}, \tag{D21}$$

which is the Eq. (18) in Sec. IV.

The biphoton probability amplitude function of the biphoton state generated by SPDC process under the same pump configuration with Fig. 1 (a) can be expressed as [32]

$$\psi(\mathbf{k}_{s\|},\mathbf{k}_{i\|}) \propto \tilde{E}_p^{(+)}(\mathbf{k}_{s\|}+\mathbf{k}_{i\|})\text{sinc}(\frac{\Delta kL}{2})e^{i\frac{\Delta kL}{2}}. \tag{D22}$$

where $\tilde{E}_p^{(+)}(\mathbf{k}_\|)$ is the slowly varying envelope of $E_p^{(+)}(\mathbf{k}_\|)$. Here, the SPDC process is assumed to occur in thick crystal with longitudinal thickness of $L$, and the term $\text{sinc}(\Delta kL/2)e^{i\Delta kL/2}$ originates from the longitudinal phase mismatch. After substituting Eq. (A2) into Eq. (D22), and restricting the signal and idler fields to the propagating fields along $z$ axis, we can obtain

$$\psi(\mathbf{k}_{s\|},\mathbf{k}_{i\|}) = C_S e^{-\frac{r_p^2|\mathbf{k}_{s\|}+\mathbf{k}_{i\|}|^2}{2}}\sin(\frac{\Delta kL}{2})e^{i\frac{\Delta kL}{2}}U(\Omega_s,k_{s\|})U(\Omega_i,k_{s\|}). \tag{D23}$$

Taking the same process leading to Eqs. (D2), (D8) and (D9), we can get the spatial Schmit number of the SPDC process in thick crystal as

$$K_S = \frac{N_K^2}{T}\int d\mathbf{k}_{s\|}\text{sinc}^2(\Delta kL)U(\Omega_s,k_{s\|})U(\Omega_i,k_{s\|})$$
$$= \frac{3}{8}\frac{\int d\mathbf{k}_{s\|}\text{sinc}^2(\Delta kL)U(\Omega_s,k_{s\|})U(\Omega_i,k_{s\|})}{\frac{2\pi}{r_p^2}}, \tag{D22}$$

which is the Eq. (19) in Sec. IV.

### APPENDIX E: ESTIMATING THE COINCIDENCE COUNT RATE OF PHOTON PAIR GENERATION

The number of photon pairs arrived at the detectors not only depends on the generation rate of the photon pairs source, but also depends on the filtering processes between the source and detectors. In the following calculations, the filtering functions of the signal/idler frequency filters are the same to that shown in Eq. (6.1) in Sec. III, and the filtering functions [39] of the collection system can be approximated by Gaussian functions

$$F_k(\mathbf{k}_s) = e^{-\frac{k_{s\|}^2}{2k_{sv}^2\sin^2\frac{\theta}{2}}}, \qquad F_k(\mathbf{k}_i) = e^{-\frac{k_{i\|}^2}{2k_{iv}^2\sin^2\frac{\theta}{2}}}, \tag{E1}$$

where $k_{sv}=\omega_s/c$ and $k_{iv}=\omega_i/c$ are the wavenumbers of the signal and idler fields in vacuum. When the frequency filtering bandwidth is not too large, and the central angular frequencies of the signal/idler frequency filters are not far away from $\omega_{p0}$, the filtering functions in Eq. (E1) can be approximated by

$$F_k(\mathbf{k}_s) \approx e^{-\frac{k_{s\|}^2}{2k_{p0}^2\sin^2\frac{\theta_c}{2}}}, \qquad F_k(\mathbf{k}_i) \approx e^{-\frac{k_{i\|}^2}{2k_{p0}^2\sin^2\frac{\theta_c}{2}}}. \tag{E2}$$

Then, the following operators are constructed

$$a_s^\dagger(t_s,\mathbf{r}_s) = (2\pi)^{-\frac{3}{2}}\sum_{l_s=0,1}\int d\Omega_s\int d\mathbf{k}_{s\|}a_s^\dagger(\Omega_s,\mathbf{k}_{s\|},0,l_s)F_\Omega(\Omega_s)F_k(\mathbf{k}_{s\|})e^{i(\mathbf{k}_{s\|}\cdot\mathbf{r}_s-\Omega_s t_s)}, \tag{E3}$$

$$a(t_s,\mathbf{r}_s) = (2\pi)^{-\frac{3}{2}}\sum_{l_s=0,1}\int d\Omega_s\int d\mathbf{k}_{s\|}a_s(\Omega_s,\mathbf{k}_{s\|},0,l_s)F_\Omega(\Omega_s)F_k(\mathbf{k}_{s\|})e^{-i(\mathbf{k}_{s\|}\cdot\mathbf{r}_s-\Omega_s t_s)}. \tag{E4}$$

Since there is no other source of photons at the signal/idler frequencies in the model shown in Fig. 1 (a), except the photon pairs generated in SpFWM, the signal/idler photon generation rate is equal to the photon pair generation rate [2]. When both of the frequency and spatial frequency filtering functions are uniform for the signal and idler photons, we can get the coincidence count rate by $\eta_i R_s$, where $R_s$ is the single side count rate of signal photon. The expression of $R_s$ is [2]

$$R_s = \eta_s\int d\mathbf{r}_s D(\mathbf{r}_s)\langle\psi|a_s^\dagger(t_s,\mathbf{r}_s)a(t_s,\mathbf{r}_s)|\psi\rangle_{s-i}, \tag{E5}$$

where $D(\mathbf{r}_s)=1$ ($r_s<r_c$) and $D(\mathbf{r}_s)=0$ ($r_s>r_c$) are due to the assumption that the single photon detectors have circle aperture with radius of $r_c$.

Substituting the expression of $|\psi\rangle_{s-i}$ in Eqs. (2) and (3) into Eq. (E5), we can get

$$R_s = (2\pi)^{-3}|\beta|^2(\gamma P_p d)^2(\sqrt{\pi}\tau_p)^2(\pi r_p^2)^2\eta_s\int d\mathbf{r}_s D(\mathbf{r}_s)$$
$$\times\langle 0|\langle 0|\sum_{l_{s1}=0,1}\sum_{l_{i1}=0,1}\int d\Omega_{s1}\int d\Omega_{i1}\int d\mathbf{k}_{s1\|}\int d\mathbf{k}_{i1\|}G_\Omega(\Omega_{s1},\Omega_{i1})G_k(\mathbf{k}_{s1\|},\mathbf{k}_{i1\|})$$
$$\times\Phi(\varphi_{ks1},\varphi_{ki1},l_{s1},l_{i1})\Theta(\Omega_{s1},k_{s1\|},0,l_{s1})\Theta(\Omega_{i1},k_{i1\|},0,l_{i1})U(\Omega_{s1},k_{s1\|})U(\Omega_{i1},k_{i1\|})a_s(\Omega_{s1},\mathbf{k}_{s1\|},0,l_{s1})a_i(\Omega_{i1},\mathbf{k}_{i1\|},0,l_{i1})$$

$$\times \sum_{l'_s=0,1} \int d\Omega'_s \int d\mathbf{k}'_{s\|} a_s^\dagger(\Omega'_s, \mathbf{k}'_{s\|}, 0, l'_s) F_\Omega(\Omega'_s) F_k(\mathbf{k}'_{s\|}) e^{-i(\mathbf{k}'_{s\|} \cdot \mathbf{r}_s - \Omega'_s t_s)}$$

$$\times \sum_{l''_s=0,1} \int d\Omega''_s \int d\mathbf{k}''_{s\|} a_s(\Omega''_s, \mathbf{k}''_{s\|}, 0, l''_s) F_\Omega(\Omega''_s) F_k(\mathbf{k}''_{s\|}) e^{i(\mathbf{k}''_{s\|} \cdot \mathbf{r}_s - \Omega''_s t_s)}$$

$$\times \sum_{l_{s22}=0,1} \sum_{l_{i2}=0,1} \int d\Omega_{s2} \int d\Omega_{i2} \int d\mathbf{k}_{s\|2} \int d\mathbf{k}_{i\|2} G_\Omega(\Omega_{s2}, \Omega_{i2}) G_k(\mathbf{k}_{s2\|}, \mathbf{k}_{i2\|}) \Phi(\varphi_{ks2}, \varphi_{ki2}, l_{s2}, l_{i2})$$

$$\times \Theta(\Omega_{s2}, k_{s2\|}, 0, l_{s2}) \Theta(\Omega_{i2}, k_{i2\|}, 0, l_{i2}) U(\Omega_{s2}, k_{s2\|}) U(\Omega_{i2}, k_{i2\|}) a_s^\dagger(\Omega_{s2}, \mathbf{k}_{s2\|}, 0, l_{s2}) a_i^\dagger(\Omega_{i2}, \mathbf{k}_{i2\|}, 0, l_{i2}) |0\rangle|0\rangle \quad . \tag{E6}$$

With the commutations

$$[a_s(\Omega''_s, \mathbf{k}''_{s\|}, 0, l''_s), a_s^\dagger(\Omega_{s2}, \mathbf{k}_{s2\|}, 0, l_{s2})] = \delta(\Omega''_s - \Omega_{s2}) \delta(\mathbf{k}''_{s\|} - \mathbf{k}_{s2\|}) \delta_{l''_s l_{s2}},$$

$$[a_s(\Omega_{s1}, \mathbf{k}_{s1\|}, 0, l_{s1}), a_s^\dagger(\Omega'_s, \mathbf{k}'_{s\|}, 0, l'_s)] = \delta(\Omega'_s - \Omega_{s1}) \delta(\mathbf{k}'_{s\|} - \mathbf{k}_{s1\|}) \delta_{l''_s l_{s1}},$$

$$[a_i(\Omega_{i1}, \mathbf{k}_{i1\|}, 0, l_{i1}), a_i^\dagger(\Omega_{i2}, \mathbf{k}_{i2\|}, 0, l_{i2})] = \delta(\Omega_{i1} - \Omega_{i2}) \delta(\mathbf{k}_{i1\|} - \mathbf{k}_{i2\|}) \delta_{l_{i1} l_{i2}}. \tag{E7}$$

Eq. (E6) can be simplified into

$$R_s = (2\pi)^{-3} |\beta|^2 (\gamma P_p d)^2 (\sqrt{\pi}\tau_p)^2 (\pi r_p^2)^2 \eta_s \int d\mathbf{r}_s D(\mathbf{r}_s) \sum_{l_{s1}=0,1} \sum_{l_{s2}=0,1} \sum_{l_{i1}=0,1} \int d\Omega_{s1} \int d\Omega_{s2} \int d\mathbf{k}_{s1\|} \int d\mathbf{k}_{s2\|} \int d\Omega_i \int d\mathbf{k}_{i\|}$$

$$\times G_\Omega(\Omega_{s1}, \Omega_i) G_k(\mathbf{k}_{s1\|}, \mathbf{k}_{i\|}) G_\Omega(\Omega_{s2}, \Omega_i) G_k(\mathbf{k}_{s2\|}, \mathbf{k}_{i\|}) \Phi(\varphi_{ks1}, \varphi_{ki}, l_{s1}, l_i) \Phi(\varphi_{ks2}, \varphi_{ki}, l_{s2}, l_i)$$

$$\times \Theta(\Omega_{s1}, k_{s1\|}, 0, l_{s1}) \Theta(\Omega_{s1}, k_{s1\|}, 0, l_{s2}) \Theta^2(\Omega_i, k_{i\|}, 0, l_i) U(\Omega_{s1}, k_{s1\|}) U(\Omega_{s2}, k_{s2\|}) U(\Omega_i, k_{i\|})$$

$$\times F_\Omega(\Omega_{s1}) F_k(\mathbf{k}_{s1\|}) F_\Omega(\Omega_{s2}) F_k(\mathbf{k}_{s2\|}) e^{-i(\mathbf{k}_{s1\|} \cdot \mathbf{r}_s - \Omega_{s1} t_s)} e^{i(\mathbf{k}_{s2\|} \cdot \mathbf{r}_s - \Omega_{s2} t_s)} \quad . \tag{E8}$$

When $r_p$ and $\tau_p$ are relatively large, the functions $G_\Omega(\Omega_{s1}, \Omega_i) = e^{-\tau_p^2(\Omega_s^2 + \Omega_i^2)/4}$ and $G_k(\mathbf{k}_{s2\|}, \mathbf{k}_{i\|}) = e^{-r_p^2(\mathbf{k}_{s2\|} + \mathbf{k}_{i\|})^2/4}$ can be approximated by $G_\Omega(\Omega_s, \Omega_i) = 2\sqrt{\pi}\delta(\Omega_s - \Omega_i)/\tau_p$ and $G_k(\mathbf{k}_{s\|}, \mathbf{k}_{i\|}) = 4\pi\delta(\mathbf{k}_{s\|} + \mathbf{k}_{i\|})/r_p^2$, respectively. When these approximations are used, Eq. (E8) can be simplified to

$$R_s = 8\pi^3 |\beta|^2 (\gamma P_p d)^2 \eta_s \pi r_c^2 \sum_{l_{s1}=0,1} \sum_{l_{s2}=0,1} \sum_{l_{i1}=0,1} \int d\Omega_s \int d\mathbf{k}_{s\|} F_\Omega^2(\Omega_s) F_k^2(\mathbf{k}_{s\|}) \Phi(\varphi_{ks}, \varphi_{ks} + \pi, l_{s1}, l_i) \Phi(\varphi_{ks}, \varphi_{ks} + \pi, l_{s2}, l_i)$$

$$\times \Theta(\Omega_s, k_{s\|}, 0, l_{s1}) \Theta(\Omega_s, k_{s\|}, 0, l_{s2}) \Theta^2(-\Omega_s, k_{s\|}, 0, l_i) U(\Omega_s, k_{s\|}) U(\Omega_i, k_{s\|}) \tag{E9}$$

According to Eq. (4.2) and the expression of $U(\Omega_s, k_{s\|})$, the terms $\Theta(\Omega_s, k_{s\|}, 0, l_{s1})$, $\Theta(\Omega_s, k_{s\|}, 0, l_{s2})$, $\Theta^2(-\Omega_s, k_{s\|}, 0, l_i)$, $U(\Omega_s, k_{s\|})$ and $U(\Omega_i, k_{s\|})$ are all nearly unit when the bandwidths of frequency and spatial frequency filtering processes are not very large. In such case, after the expressions of $F_\Omega(\Omega_s)$ and $F_k(\mathbf{k}_{s\|})$ in Eq. (6.1) and Eq. (E2) are substituted into Eq. (9), this equation can be simplified to

$$R_s = 8\pi^3 |\beta|^2 (\gamma P_p d)^2 \eta_s \pi r_c^2 \sum_{l_{s1}=0,1} \sum_{l_{s2}=0,1} \sum_{l_{i1}=0,1} \int d\Omega_s \int d\mathbf{k}_{s\|} F_\Omega^2(\Omega_s) F_k^2(\mathbf{k}_{s\|}) \Phi(\varphi_{ks}, \varphi_{ks} + \pi, l_{s1}, l_i) \Phi(\varphi_{ks}, \varphi_{ks} + \pi, l_{s2}, l_i)$$

$$= \frac{\pi\sqrt{\pi} n_{p0}^2}{9 n_s n_i} (\gamma P_p d)^2 \eta_s \Omega_c (k_{p0} r_c \sin\frac{\theta_c}{2})^2 [(\frac{1+r_\chi}{2})^2 + (\frac{1-r_\chi}{2})^2]. \tag{E10}$$

With Eq. (E10), the coincidence count $n_c$ per pump pulse can be obtained as $n_c = \eta_i(\sqrt{\pi}\tau_p) R_s$, which is the Eq. (20) in Sec. V.

**APPENDIX F: PARAMETERS FOR ESTIMATING THE COINCIDENCE COUNT RATE**

Firstly, we should determine the film thickness within which the material loss can be neglected for the SpFWM process, since a precondition of the biphoton state in Eq. (1) is the assumption of negligible material loss. Otherwise, we cannot directly construct the Schrödinger equation by the Hamiltonian in Eq. (A9) [40]. Table F1 shows the pump and signal wavelengths $\lambda_p$ and $\lambda_s$ used in the calculations, as well as the absorption coefficients $\alpha_p$, $\alpha_s$ and $\alpha_I$ at the pump, signal and idler wavelengths, respectively. For $Bi_2Se_3$, the absorption coefficients $\alpha_p=1.885\times10^5$/cm, $\alpha_s=1.057\times10^5$/cm, and $\alpha_i=3.161\times10^5$/cm at 800.0nm, 980.0nm, and 675.8nm, respectively [41], while the three values for graphene are $\alpha_p \approx \alpha_s \approx \alpha_i = 5.64\times10^5$/cm when the pump, signal and idler wavelengths are 1550nm, 1530nm, and 1570.5nm [42]. Moreover, because the power used in estimating the coincidence counts is not low, and we should determine whether the nonlinear absorptions of the pump light are considerable, including the two photon absorption (TPA) and saturation absorption (SA). In $Bi_2Se_3$, according to Ref. [34], TPA does not occur even when the incident irradiance at 800nm is $I_{sat}=10.4$GW/cm$^2$, which corresponds to $P_p=8.2$kW with $r_p=5\mu$m. Moreover, the SA threshold of $Bi_2Se_3$ at 800nm is $I_{sat}=10.12$GW/cm$^2$ [34] and corresponds to $P_p=7.9$kW with $r_p=5\mu$m, which is much higher than the power used in the calculations about $Bi_2Se_3$ in Sec. V. In graphene, the TPA coefficient and SA threshold of light with wavelength around 1.5$\mu$m are $\alpha_T=0.9\times10^4$cm/GW and $I_{sat}=3$GW/cm$^2$, respectively [43]. With the pump powers and $r_p=5\mu$m used in the calculations about graphene in Fig.5,

$\alpha_T P_p/A_{eff}$ is much less than $\alpha_p$ of graphene and the pump light intensity $P_p/A_{eff}$ is lower than $I_{sat}$, where $A_{eff} = \pi r_p^2$ is the area of pump light. With these comparisons, we can conclude that the TPA and SA effects can be neglected in the calculations for both the $Bi_2Se_3$ and graphene in Fig.5. When both the linear absorption and reflection are considered, the material losses of single-layered $Bi_2Se_3$ with thickness $d_L$=1nm [44] are 0.18dB, 0.10dB, and 0.31dB for the incident pump light, the generated signal and idler fields, respectively. In similar way, we can obtain that the material losses at the pump, signal and idler fields are all around 0.28dB, for four-layered graphene with single layer thickness $d_L$=0.3nm [45]. These loss values are all small enough to be neglected, and therefore the thicknesses $d=N_L d_L$ can be directly substituted into Eq. (20) to perform the calculations for single/four-layered $Bi_2Se_3$ and graphene.

Secondly, other parameters used in estimating the coincidence count rate $n_c$ per pulse in different materials are listed in the following table. Both $Bi_2Se_3$ and graphene belong to hexagonal crystal classes, and therefore $\chi^{(3)}_{xxxx}=\chi^{(3)}_{yyxx}+\chi^{(3)}_{xyyx}+\chi^{(3)}_{xyxy}$ is satisfied [46]. As far as we know, there is no reports on the ratios of $\chi^{(3)}_{yyxx}$ to $\chi^{(3)}_{xyyx}$ and $\chi^{(3)}_{xyxy}$. In Table F1, we assume the ratios are both 1, i.e., $r_\chi$=1/3. According to Eq. (20), $n_c$ per pulse will get its maximum (minimum) at $r_\chi$=1 ($r_\chi$=0) when $r_\chi$ is varied and other parameters are fixed. It can be easily obtained that this maximum (minimum) is only 1.8 (0.9) times higher (lower) than $n_c$ per pulse when $r_\chi$=1/3, i.e., they are of the same order of magnitude. Therefore, the calculated $n_c$ per pulse at $r_\chi$=1/3 is a reasonable estimation of the value in real experiment.

Table F1 Some parameters used in the calculations for Fig. 5

| $Bi_2Se_3$ | Graphene |
|---|---|
| $d_L$=1nm | $d_L$=0.3nm |
| $\lambda_p$=800nm | $\lambda_p$=1530nm |
| $\lambda_s$=980nm | $\lambda_s$=1550nm |
| $\Omega_c/2\pi$=125GHz | $\Omega_c/2\pi$=125GHz |
| $\theta_c = 6°$ | $\theta_c = 6°$ |
| $r_c$=1mm | $r_c$=1mm |
| $n_2$=2.26×10$^{-14}$m$^2$/W [34] | $\chi^{(3)}_{xxxx}=10^{-15} m^2/V^2$ [33] |
| $n_{p0}$=3.8, $n_s$=3, $n_i$=4.6 [41] | |
| $\alpha_p$=1.885×10$^5$/cm [41] | $n_{p0}\approx n_{s,i}$=3.2 [42] |
| $\alpha_s$=1.057×10$^5$/cm [41] | $\alpha_p\approx\alpha_s\approx\alpha_i$=5.64×10$^5$/cm [40] |
| $\alpha_i$=3.161×10$^5$/cm [41] | $r_\chi$=1/3 |
| $r_\chi$=1/3 | |